\documentclass[final,twocolumn]{IEEEtran}

\usepackage{graphicx}
\usepackage{amssymb}
 
\usepackage{amsmath}
\usepackage{dsfont}
\usepackage{bm}
\usepackage{eucal}
\usepackage{cite}
\usepackage{color}
\usepackage{amsthm}
\usepackage{algorithmic}
\usepackage{algorithm}
 \usepackage{subcaption}

\DeclareMathOperator{\sinc}{sinc}

\def\C{{\mathds C}}

\newcommand{\be}{\begin{equation}}
\newcommand{\ee}{\end{equation}}

\newcommand{\bzero}{{\mbox{\boldmath $0$}}}
\newcommand{\bI}{{\mbox{\boldmath $I$}}}
\newcommand{\bz}{{\mbox{\boldmath $z$}}}

\newcommand{\bx}{{\mbox{\boldmath $x$}}}

\newcommand{\bbeta}{{\mbox{\boldmath $\beta$}}}
\newcommand{\bB}{{\mbox{\boldmath $B$}}}

\newcommand{\bG}{{\mbox{\boldmath $G$}}}

\newcommand{\bL}{{\mbox{\boldmath $L$}}}

\newcommand{\bZ}{{\mbox{\boldmath $Z$}}}
\newcommand{\bR}{{\mbox{\boldmath $R$}}}
\newcommand{\bV}{{\mbox{\boldmath $V$}}}

\newcommand{\bX}{{\mbox{\boldmath $X$}}}
\newcommand{\bS}{{\mbox{\boldmath $S$}}}

\newcommand{\bP}{{\mbox{\boldmath $P$}}}

\newcommand{\bH}{{\mbox{\boldmath $H$}}}

\newcommand{\tr}{\mbox{\rm Tr}\, }

\newcommand{\bSigma}{\mbox{\boldmath{$\Sigma$}}}
\newcommand{\bU}{{\mbox{\boldmath $U$}}}

\newcommand{\bM}{{\mbox{\boldmath $M$}}}
\newcommand{\diag}{\mbox{diag}\, }

\newcommand{\bT}{{\mbox{\boldmath $T$}}}

\newcommand{\ds}{\displaystyle}

\newcommand{\test}{\mbox{$
\begin{array}{c}
\stackrel{ \stackrel{\textstyle H_1}{\textstyle >} }{ 
\stackrel{\textstyle <}{ \textstyle  H_0} }

\end{array}
$}}

\def\cC{\mbox{$\CMcal C$}}
\def\cN{\mbox{$\CMcal N$}}

\begin{document}

\title{A Unified Theory of Adaptive Subspace Detection. Part II: Numerical Examples}

\author{Pia Addabbo, \IEEEmembership{Senior Member, IEEE},  
Danilo Orlando, \IEEEmembership{Senior Member, IEEE},  
Giuseppe Ricci$^{*}$, \IEEEmembership{Senior Member, IEEE}, 
and Louis L. Scharf, \IEEEmembership{Life Fellow, IEEE}
\thanks{Pia Addabbo is with Dipartimento di Ingegneria, Università degli Studi del Sannio, Piazza Roma 21, 82100 Benevento, Italy. E-mail: {\tt paddabbo@unisannio.it}.}
\thanks{Danilo Orlando is with Università degli Studi “Niccolò Cusano”, via
Don Carlo Gnocchi, 3, 00166 Roma, Italy. E-mail: {\tt danilo.orlando@unicusano.it}.}
\thanks{Giuseppe Ricci is with the Dipartimento di Ingegneria dell'Innovazione,
Universit\`{a} del Salento, Via Monteroni, 73100 Lecce, Italy.
E-Mail: {\tt giuseppe.ricci@unisalento.it}.}
\thanks{Louis L. Scharf is with the Departments of Mathematics and Statistics, Colorado
State University, Fort Collins, CO, USA
E-Mail: {\tt scharf@colostate.edu}. His work is supported by the US Office of Naval Research under 
contract N00014-21-1-2145, and by the US Airforce Office of Scientific Research under contract AF 9550-21-1-0169.}
\thanks{$^{*}$Corresponding author}         
}

\maketitle

\begin{abstract}
This paper is devoted to the performance analysis of the detectors proposed in the companion paper
\cite{SubSpFrame_PART_I} where a comprehensive design framework is presented for the
adaptive detection of subspace signals. 
The framework addresses four variations on subspace detection: 
the subspace may be known or 
known only by its dimension; consecutive visits to the subspace may be unconstrained
or they may be constrained by a prior  probability distribution. In this paper, Monte Carlo simulations are used to compare 
the generalized likelihood ratio (GLR) detectors 
derived in \cite{SubSpFrame_PART_I}   
with  estimate-and-plug (EP) approximations of the GLR detectors.
Remarkably, the EP approximations
appear here for the first time (at least to the best of the authors' knowledge).
The numerical examples indicate that GLR detectors
are effective for the detection of {\em partially-known signals} 
affected by inherent uncertainties due to the system or the operating environment. In particular, 
if the signal subspace is known, GLR detectors tend to ouperform EP
detectors. If, instead, the signal subspace is 
known only by its dimension, the performance of GLR and EP detectors is very similar.
\end{abstract}


\section{Introduction and Problem Formulation}
Adaptive detection of targets modeled as belonging to suitable subspaces has been widely investigated by the
signal processing community with applications ranging from radar 
and sonar to communications and hyperspectral imaging
\cite{BDMGR,BBORS2007,Scharf-Friedlander1994,scharfRevewSubspace,hyperSubspace,8850120,9387095,BCCRV}.
In the context of radar signal processing, the general framework devised in \cite{Kelly-Forsythe}
for homogeneous environments where  
test and training samples share 
the same Gaussian distribution 
has been extended over the years by including
unknown scaling differences between test and training 
samples \cite{Kraut-Scharf1999}, structured 
interference components as
well as non-Gaussian disturbances  \cite{subspaceNonGaussian,Gini-Farina2002}.


As stated in the companion paper \cite{SubSpFrame_PART_I}, most of these works deal with deterministic targets 
embedded in random disturbance with unknown covariance matrix. The term deterministic means that
target signatures do not obey any prior distribution and, hence, target coordinates within
the subspace are not random variables. Generally speaking, this design assumption
is referred to as {\em first-order (signal) model}. 
On the contrary in a  
{\em second-order (signal) model}, 
the signal coordinates in the subspace are random variables and parameters of the signal signature appear in second-order statistics such as the covariance matrix.
The first application of the second-order model to target detection 
in partially-homogeneous Gaussian environment can be found in \cite{ASAP04}, where the estimate-and-plug  (EP) approximation to the
generalized likelihood ratio test (GLRT) has been used \cite{Robey}. This approach
consists in computing the GLRT assuming that a subset of parameters is known. Then,
in order to make the detector fully adaptive, the known parameters 
are replaced with suitable estimates.
The main advantage of the estimate-and-plug approximation is 
that the resulting detectors have 
lower computational complexity 
than their GLR counterparts. But there is generally a loss in performance, and it is this loss that we aim to quantify in this paper.

The second-order model has been further investigated in the companion paper \cite{SubSpFrame_PART_I}, where
a unified theoretical framework for subspace adaptive detection (including the first-order model)
in Gaussian disturbance has been
devised. More importantly, the exact GLRT or suitable approximations of it 
have been therein derived for the first time
(at least to the best of authors' knowledge). 
These approximations rely on cyclic estimation procedures \cite{Stoica_alternating}
where, at each step, closed-form updates of the parameter estimates are computed.

Following the 
conventions of \cite{SubSpFrame_PART_I},\footnote{
{\em Notation:} in the sequel, vectors and matrices are denoted by boldface lower-case and upper-case letters, respectively.
Symbols $\det(\cdot)$, $\tr(\cdot)$, 
$(\cdot)^T$, 
and $(\cdot)^\dag$ denote the determinant, trace, 
transpose, 
and conjugate transpose, respectively. 
As to numerical sets, 
$\C$ is the set of 
complex numbers, $\C^{N\times M}$ is the Euclidean space of $(N\times M)$-dimensional 
complex matrices, and $\C^{N}$ is the Euclidean space of $N$-dimensional 
complex vectors. 
$\bI_n$ and $\bzero_{m,n}$ stand for the $n \times n$ identity matrix and the $m \times n$ null matrix.
$\langle \bH \rangle$ denotes the space spanned by the columns of the matrix $\bH
\in \C^{N\times M}$.
Given $a_1, \ldots, a_N \in\C$, 
$\diag(a_1, \ldots, a_N)
\in\C^{N\times N}$ indicates 
the diagonal matrix whose $i$th diagonal element is $a_i$.
We write $\bz\sim \cC\cN_N(\bx, \bSigma)$ to say that the $N$-dimensional random vector 
$\bz$ is  a complex normal random vector with mean vector $\bx$ and covariance matrix $\bSigma$. 
Moreover, $\bZ=[\bz_1 \cdots \bz_K] \sim \cC\cN_{NK}(\bX, \bI_K\otimes \bSigma)$, with $\otimes$ denoting Kronecker product and
$\bX=[\bx_1 \cdots \bx_K]$,
means that
$\bz_k\sim \cC\cN_N(\bx_k, \bSigma)$ and the columns of $\bZ$ are statistically independent.
The acronym PDF stands for probability density function.
$\widehat{\bR}_i$ and $\widehat{\gamma}_i$ will denote the (possibly approximated) maximum likelihood (ML) estimates of $\bR$ and $\gamma$, respectively, under the $H_i$ hypothesis, $i=0,1$.
}
let us consider a detection system that collects data from a primary and a secondary channel.
Data under test are those from the primary channel and are denoted by
$\bZ_P=[\bz_1 \cdots \bz_{K_P}]\in\C^{N\times K_P}$, whereas data from the secondary channel,
used for the estimation of the disturbance parameters, are indicated by 
$\bZ_S=[\bz_{K_P+1} \cdots \bz_{K_P+K_S}]\in\C^{N\times K_S}$.
In the case of first-order models, the detection problem at hand can be formulated as \cite{SubSpFrame_PART_I}
\begin{equation}
\left\{
\begin{array}{ll}
H_{0}: & 
\left\{
\begin{array}{l}
\bZ_P \sim \cC \cN_{NK_P} (\bzero_{N,K_P}, \bI_{K_P}\otimes \bR) \\ \bZ_S \sim \cC\cN_{NK_S} (\bzero_{N,K_S},  \bI_{K_S}\otimes \gamma\bR)
\end{array} \right. \\ \\
H_{1}: &  
\left\{
\begin{array}{l}
\bZ_{P} \sim \cC\cN_{NK_P} (\bH \bX, \bI_{K_P}\otimes\bR) 
\\ 
\bZ_S \sim {\cC\cN}_{NK_S} (\bzero_{N,K_S}, \bI_{K_S}\otimes\gamma \bR)
\end{array} \right.
\end{array} 
\right.
\label{FO-HT}
\end{equation}
where 
$\bH \in \C^{N \times r}$ is either a known matrix or 
an unknown matrix with known rank $r$, $r \leq N$, $\bX=[\bx_1 \cdots \bx_{K_P}] \in \C^{r \times K_P}$ is the matrix of the unknown signal coordinates,  
$\bR \in \C^{N \times N}$ is an unknown positive definite
covariance matrix while $\gamma>0$ is either a known or an unknown parameter. 
In the following, we suppose that $K_S \geq N$. Without loss of generality, we assume that 
$\bH$ is an arbitrary unitary basis for a subspace that is either known or known only by its dimension.

The hypothesis test based upon the second-order model is formulated as 
\begin{equation}
\left\{\!\!\!
\begin{array}{ll}
H_{0}: & \!\!\! \!\!
\left\{\!
\begin{array}{l}
\bZ_P \sim \cC \cN_{NK_P} (\bzero_{N,K_P}, \bI_{K_P}\otimes \bR) \\ \bZ_S \sim \cC\cN_{NK_S} (\bzero_{N,K_S},  \bI_{K_S}\otimes \gamma\bR) 
\end{array} \right.
\\ \\
H_{1}: & \!\!\!\!\! 
\left\{\!
\begin{array}{l}
\bZ_{P} \sim \cC\cN_{NK_P} (\bzero_{N,K_P}, \bI_{K_P}\otimes (\bH \bR_{xx} \bH^{\dag}+\bR)) \\
\bZ_S \sim {\cC\cN}_{NK_S} (\bzero_{N,K_S}, \bI_{K_S}\otimes\gamma \bR)
\end{array} \right.
\end{array} 
\right.
\label{SO-HT}
\end{equation}
where $\bR_{xx} \in \C^{r \times r}$ is an unknown positive semidefinite 
matrix (in order to account for possible correlated sources \cite[and references therein]{9309189}).
It is important to observe that when the scaling factor $\gamma$ is known, both \eqref{FO-HT} and \eqref{SO-HT}
account for a homogeneous environment where primary and secondary data
share the same statistical characterization of the disturbance. 
In fact, secondary data can be equalized, so it is as if   $\gamma=1$.
On the other hand, when such a parameter is unknown, the corresponding operating scenario
is referred to as partially-homogeneous \cite{Kraut-Scharf1999}.
The latter model is an extension of the homogeneous environment and, 
though keeping a relative mathematical tractability, it leads to an increased robustness to inhomogeneities since the
assumed difference in power level accounts for terrain type
variations, height profile, and shadowing which may appear in practice \cite{Ward1994}.

In this paper, 
we assess the performance of the GLR detectors
derived in the first part \cite{SubSpFrame_PART_I} by analyzing probability of detection 
and false alarm rate. In addition, we compare these performance metrics with those returned by the estimate-and-plug approximations (that are devised in the next
subsections). Even though these competitors
can be obtained by exploiting existing derivations \cite{BDMGR,Bresler,ASAP04}, some of them appear here for the first time. 

The remainder of this paper is organized as follows. In the next section, the detection
architectures devised in the first part \cite{SubSpFrame_PART_I} are summarized and the expressions of the
estimate-and-plug competitors are given. In Section \ref{sec:Performance}, the performance
of the GLR and EP detectors are investigated and discussed through numerical examples.
Section \ref{sec:Conclusions} contains concluding remarks and future research tracks.

\section{Detection Architectures}
\label{sec:Architectures}
The aim of this section is twofold. First, in order to make this second part self-contained, 
we briefly summarize the decision schemes developed in the companion paper. Second,
we provide the expressions of the competitors that are based upon the 
estimate-and-plug paradigm \cite{Robey,BOR_MC2009}.
Recall that this approach consists in computing the GLRT under the assumption that some parameters are known and in 
replacing them with suitable estimates. For the case at hand, the covariance matrix of the disturbance is initially supposed 
known and in the final decision statistic it is replaced by the sample 
covariance matrix (SCM) computed from secondary data only.

\subsection{GLRT-based Detectors Summary}
The detectors described in this subsection are those derived in the first part of this work \cite{SubSpFrame_PART_I}.
Throughout, the log-likelihood function under $H_i$ is denoted by $L_i(\cdot)$, $i=0,1$.

\subsubsection{First-order models}
Consider problem \eqref{FO-HT}, the related four cases are listed 
below.\footnote{As in the companion paper \cite{SubSpFrame_PART_I}, the generic detection threshold will be indicated by $\eta$.}
\begin{itemize}
\item {\bf Known subspace $\langle \bH \rangle$, known $\gamma$}:
the GLRT for problem \eqref{FO-HT} with $\gamma=1$ is referred to as a 
first-order detector for a signal in a known subspace in a homogeneous environment (FO-KS-HE) and is given by
\be
\label{eq:1S-FO-GLRT-KH-HE}
\frac{{\det}\left[ \bI_{K_P} + \bZ_P^{\dag}
\bS_S^{-1} \bZ_P \right]}{
{\det} \left[ \bI_{K_P} +
\left( \bS_S^{-1/2} \bZ_P \right)^{\dag} \bP_G^{\perp}
\left( \bS_S^{-1/2} \bZ_P \right)
\right]
}
\test \eta
\ee
where $\bS_S=\bZ_S\bZ_S^\dag$ and $\bP_G^{\perp}=\bI_N-\bP_G$ with
$\bP_G=\bG(\bG^\dag\bG)^{-1}\bG^\dag$ and
$\bG=\bS_S^{-1/2}\bH$.
\item {\bf Known subspace $\langle \bH \rangle$, unknown $\gamma$}:
under the assumption $r <N$ and $\min(K_P,N-r) > \frac{NK_P}{K}$, the GLRT for problem \eqref{FO-HT} with $\gamma>0$ is referred to as a 
first-order detector for a signal in a known subspace in a partially-homogeneous environment 
(FO-KS-PHE), and is given by
\be
\frac{\widehat{\gamma}_0^{\frac{K_P(K-N)}{K}} {\det}\left[ \frac{1}{\widehat{\gamma}_0}  \bI_{K_P} + \bM_0 \right]
}{\widehat{\gamma}_1^{\frac{K_P(K-N)}{K}}
{\det} \left[ \frac{1}{\widehat{\gamma}_1}  \bI_{K_P} +
\bM_1
\right]
}
\test \eta
\label{eq:1S-FO-GLRT-KH-PHE}
\ee
where $\bM_0=\bZ_P^{\dag}\bS_S^{-1} \bZ_P$, 
$\bM_1=\left( \bS_S^{-1/2} \bZ_P \right)^{\dag} \bP_G^{\perp} \left( \bS_S^{-1/2} \bZ_P \right)$,
and $\widehat{\gamma}_i$, $i=0,1,$ can be computed using Theorem 1 of \cite{SubSpFrame_PART_I}.
\item {\bf Unknown subspace $\langle \bH \rangle$, known $\gamma$}: in this case,
if $\min(N,K_P) \geq r +1$, the GLRT for problem \eqref{FO-HT} with $\gamma=1$ is referred to as a 
first-order detector for a signal in an unknown subspace in a homogeneous environment (FO-US-HE), and is given by
\be
\prod_{i=N-r+1}^{N} \left( 1 + \sigma^2_i
 \right)
\test \eta
\label{eq:1S-FO-GLRT-UH-HE}
\ee
where $\sigma^2_1\leq\ldots\leq\sigma^2_N$ are the eigenvalues
of $\bS_S^{-1/2}\bZ_P\bZ_P^\dag\bS_S^{-1/2}$. When $\min(N,K_P)< r +1$, the GLRT reduces to
\be
{\det}\left( \bI_N + \bS_S^{-1/2} \bZ_P \bZ_P^{\dag}
 \bS_S^{-1/2}\right)
\test \eta.
\label{eq:1S-FO-GLRT-UH-HE-r=N}
\ee
\item {\bf Unknown subspace $\langle \bH \rangle$, unknown $\gamma$}:
under the conditions $\min(N,K_P) \geq r+1$ and $\min(N,K_P) > {NK_P}/{K}+r$, the GLRT for problem \eqref{FO-HT} is referred to as 
a first-order detector for a signal in an unknown subspace in a partially-homogeneous environment 
(FO-US-PHE), and is given by 
\be
\frac{\widehat{\gamma}_0^{N\left(1-\frac{K_P}{K}\right)} 
\prod_{i=1}^{N} \left( \frac{1}{\widehat{\gamma}_0} + \sigma^2_i  \right)
}{
\widehat{\gamma}_1^{N\left(1-\frac{K_P}{K}\right)-r} 
\prod_{i=1}^{N-r} \left( \frac{1}{\widehat{\gamma}_1} + \sigma^2_i  \right)
}
\test \eta
\label{eq:1S-FO-GLRT-UH-PHE}
\ee
where $\widehat{\gamma}_0$ and $\widehat{\gamma}_1$ are computed using 
Corollary 2 and 1 of \cite{SubSpFrame_PART_I}, respectively.
\end{itemize}

\subsubsection{Second-order models}
As for problem \eqref{SO-HT}, the expressions of the related decision rules are summarized below.
\begin{itemize}
\item {\bf Known subspace $\langle \bH \rangle$, known $\gamma$}:
the approximate GLRT  for problem (\ref{SO-HT})
is referred to as a 
second-order detector for a signal in a known subspace in a homogeneous environment 
(SO-KS-HE), and
is  given by
\be
L_1(\widehat{\bR}_1, \widehat{{\bR}}_{xx},\bH, 1; \bZ)
-L_0(\widehat{\bR}_0,  1; \bZ)
\test \eta
\label{eq:1S-SO-GLRT-UK-HE}
\ee
where $L_0(\widehat{\bR}_0,  1; \bZ)$ is the logarithm of (5) in \cite{SubSpFrame_PART_I} with $\gamma=1$, while $L_1(\widehat{\bR}_1, \widehat{{\bR}}_{xx},\bH, 1; \bZ)$ is given by eq. (36) of \cite{SubSpFrame_PART_I}, with
$\widehat{\bR}_{xx}$ and $\widehat{\bR}_1$ obtained by iterating equations (39) and (40) of \cite{SubSpFrame_PART_I}
(the procedure is summarized in Algorithm \ref{alg:alternating-SO-KS-HE})
until the following convergence criterion is not satisfied:
$\Delta L_1=|L_1(\bR^{(n)}, {\bR}_{xx}^{(n)},\bH, 1; \bZ)
-L_1(\bR^{(n-1)}\!\!\!\!, 
{\bR}_{xx}^{(n-1)}\!\!\!\!,\bH, 1; \bZ)|
/|L_1(\bR^{(n-1)}\!\!\!\!, {\bR}_{xx}^{(n-1)}\!\!\!\!,\bH, 1; \bZ)|$ $
\leq \epsilon_1$ with $\epsilon_1>0$.
\item {\bf Known subspace $\langle \bH \rangle$, unknown $\gamma$}: in this case, 
an approximation of the GLRT for problem \eqref{SO-HT} is referred to as a 
second-order detector for a signal in a known subspace in a partially-homogeneous environment
(SO-KS-PHE), and is given by
\be
L_1(\widehat{\bR}_1, \widehat{\bR}_{xx}, \bH, \widehat{\gamma}_1; \bZ)-L_0(\widehat{\bR}_0,  \widehat{\gamma}_0; \bZ)
\test \eta
\label{eq:1S-SO-GLRT-KH-PHE}
\ee
where
$L_0(\widehat{\bR}_0,\widehat{\gamma}_0; \bZ)$ is the logarithm of 
the maximum of (5) in \cite{SubSpFrame_PART_I} with respect to $\gamma$
obtained by using Theorem 1 of \cite{SubSpFrame_PART_I}, while
$\widehat{\bR}_1$, $\widehat{\bR}_{xx}$, and $\widehat{\gamma}_1$ are computed through
the alternating estimation procedure exploiting (40) and (39) of \cite{SubSpFrame_PART_I}
in conjunction with Theorem 6 of \cite{SubSpFrame_PART_I}.
Again, the procedure, summarized in Algorithm \ref{alg:alternating-SO-KS-PHE},
terminates when the following condition is true: $\Delta L_2=|L_1(\bR^{(n)}, {\bR}_{xx}^{(n)},\bH, \gamma^{(n)}; \bZ)
-L_1(\bR^{(n-1)}\!\!\!\!, 
{\bR}_{xx}^{(n-1)}\!\!\!\!,\bH, \gamma^{(n-1)}; \bZ)|
/|L_1(\bR^{(n-1)}\!\!\!\!, {\bR}_{xx}^{(n-1)}\!\!\!\!,\bH, $ $ \gamma^{(n-1)}; \bZ)|
\leq \epsilon_2$ with $\epsilon_2>0$.
\item {\bf Unknown subspace $\langle \bH \rangle$, known $\gamma$}:
let $\tilde{\bR}_{xx}=\bH\bR_{xx}\bH^\dag$, then,
the GLRT for problem \eqref{SO-HT}
is referred to as a 
second-order detector for a signal in an unknown subspace in a homogeneous environment 
(SO-US-HE), and
is  given by
\be
L_1(\widehat{\bR}_1, \widehat{\tilde{\bR}}_{xx}, 1; \bZ)
-L_0(\widehat{\bR}_0,  1; \bZ)
\test \eta
\label{eq:1S-SO-GLRT-UK-HE}
\ee
where $L_0(\widehat{\bR}_0,1;\bZ)$ is given by the logarithm of (5) in \cite{SubSpFrame_PART_I}
and the expression of $L_1(\widehat{\bR}_1, \widehat{\tilde{\bR}}_{xx}, 1; \bZ)$ can be found
exploiting Theorem 3 of \cite{SubSpFrame_PART_I} with $\gamma=1$.
\item {\bf Unknown subspace $\langle \bH \rangle$, unknown $\gamma$}:
the GLRT for problem \eqref{SO-HT} is referred to as a
second-order detector for a signal in an unknown subspace in a partially-homogeneous environment
(SO-US-PHE), and is given by
\be
L_1(\widehat{\bR}_1, \widehat{\tilde{\bR}}_{xx}, \widehat{\gamma}_1; \bZ)-L_0(\widehat{\bR}_0,  \widehat{\gamma}_0; \bZ)
\test \eta
\label{eq:1S-SO-GLRT-UH-PHE}
\ee
where $L_0(\widehat{\bR}, \widehat{\gamma}_0; \bZ)$ is the logarithm of the maximum of (5) in \cite{SubSpFrame_PART_I} 
with respect to $\gamma$ obtained by 
using Theorem 1 of \cite{SubSpFrame_PART_I} and 
$L_1(\widehat{\bR}_1, \widehat{\tilde{\bR}}_{xx}, \widehat{\gamma}_1; \bZ)$ is computed
by jointly exploiting Theorems 3 and 5 of \cite{SubSpFrame_PART_I}.
\end{itemize}

\begin{algorithm}[tb!]
\caption{Alternating procedure for SO-KS-HE}
\label{alg:alternating-SO-KS-HE}
\begin{algorithmic}[1]
\REQUIRE $\epsilon_1$, $\bbeta^{(0)}$ 
\ENSURE $\widehat{\bR}_{xx}$, $\widehat{\bR}_1$
\STATE Set $n=0$ 
\STATE Estimate $\tilde{\bR}_{1.2}^{(n+1)}$ and $\bR_{xx}^{(n+1)}$, given $\bbeta^{(n)}$ using eq. (39) of \cite{SubSpFrame_PART_I}
\STATE Estimate $\bbeta^{(n+1)}$ given $\tilde{\bR}_{1.2}^{(n+1)}$ and $ \bR_{xx}^{(n+1)}$ by eq. (40) of \cite{SubSpFrame_PART_I}
\STATE Set $n=n+1$
\STATE If $\Delta L_1 \leq \epsilon_1$ go to step 6 else go to step 2
\STATE \textbf{Output:} $\widehat{\bR}_{xx}$, $\widehat{\bR}_1$ computed using $\bbeta^{(n)}$, $\tilde{\bR}_{1.2}^{(n)}$ and $\bR_{xx}^{(n)}$
\end{algorithmic}
\end{algorithm}
\begin{algorithm}[tb!]
\caption{Alternating procedure for SO-KS-PHE}
\label{alg:alternating-SO-KS-PHE}
\begin{algorithmic}[1]
\REQUIRE $\epsilon_2$, $\bbeta^{(0)}$ 
\ENSURE $\widehat{\bR}_{xx}$, $\widehat{\bR}_1$,
$\widehat{\gamma}_1$
\STATE Set $n=0$ 
\STATE Estimate $\tilde{\bR}_{1.2}^{(n+1)}$, $\bR_{xx}^{(n+1)}$, and $\gamma^{(n+1)}$, given $\bbeta^{(n)}$ 
using eq. (39) and {\em Theorem 6} of \cite{SubSpFrame_PART_I}
\STATE Estimate $\bbeta^{(n+1)}$ given $\tilde{\bR}_{1.2}^{(n+1)}$, $\bR_{xx}^{(n+1)}$, and $\gamma^{(n+1)}$
by eq. (40) of \cite{SubSpFrame_PART_I}
\STATE Set $n=n+1$
\STATE If $\Delta L_2 \leq \epsilon_2$ go to step 6 else go to step 2
\STATE \textbf{Output:} $\widehat{\bR}_{xx}$, $\widehat{\bR}_1$, and $\widehat{\gamma}_1$ computed using $\bbeta^{(n)}$, $\tilde{\bR}_{1.2}^{(n)}$, $\bR_{xx}^{(n)}$, and $\gamma^{(n)}$ 
\end{algorithmic}
\end{algorithm}

\bigskip

For the reader ease, we summarize the steps required to compute the GLRs of all of these detectors 
in Algorithms \ref{alg:FO-KS-HE}-\ref{alg:SO-US-PHE}.

\begin{algorithm}[tb!]
\caption{FO-KS-HE}
\label{alg:FO-KS-HE}
\begin{algorithmic}[1]
\REQUIRE $\bZ_P, \bZ_S, \bH$
\ENSURE Decision statistic of FO-KS-HE
\STATE Compute $\bS_S^{-1/2} = (\bZ_S \bZ_S^\dag)^{-1/2}$
\STATE Compute  $\bG=\bS_S^{-1/2}\bH$
\STATE Compute $\bP_G^{\perp}=\bI_N-\bG(\bG^\dag\bG)^{-1}\bG^\dag$ 
\STATE \textbf{Output:} $\frac{{\det}\left[ \bI_{K_P} + \bZ_P^{\dag}\bS_S^{-1} \bZ_P \right]}{{\det} \left[ \bI_{K_P} + \left( \bS_S^{-1/2} \bZ_P \right)^{\dag} \bP_G^{\perp}\left( \bS_S^{-1/2} \bZ_P \right)\right]}$
\end{algorithmic}
\end{algorithm}
\begin{algorithm}[tb!]
\caption{FO-KS-PHE}
\label{alg:FO-KS-PHE}
\begin{algorithmic}[1]
\REQUIRE $\bZ_P, \bZ_S, \bH$
\ENSURE Decision statistic of FO-KS-PHE
\STATE If $\min(K_P,N-r) > \frac{NK_P}{K}$ go to step 2 else end
\STATE Compute $\bS_S^{-1/2} = (\bZ_S \bZ_S^\dag)^{-1/2}$
\STATE Compute $\bS_S^{-1/2}\bZ_P$
\STATE Compute $\bM_0=\bZ_P^{\dag}\bS_S^{-1} \bZ_P$
\STATE Compute $\widehat{\gamma}_0$, using Theorem 1 of \cite{SubSpFrame_PART_I}
\STATE Compute  $\bG=\bS_S^{-1/2}\bH$
\STATE Compute $\bP_G^{\perp}=\bI_N-\bG(\bG^\dag\bG)^{-1}\bG^\dag$ 
\STATE Compute $\bM_1=\left( \bS_S^{-1/2} \bZ_P \right)^{\dag} \bP_G^{\perp} \left( \bS_S^{-1/2} \bZ_P \right)$
\STATE Compute $\widehat{\gamma}_1$, using Theorem 1 of \cite{SubSpFrame_PART_I}
\STATE \textbf{Output:} $\frac{\widehat{\gamma}_0^{\frac{K_P(K-N)}{K}} {\det}\left[ \frac{1}{\widehat{\gamma}_0}  \bI_{K_P} + \bM_0 \right]}{\widehat{\gamma}_1^{\frac{K_P(K-N)}{K}}
{\det} \left[ \frac{1}{\widehat{\gamma}_1}  \bI_{K_P} +\bM_1\right]}$
\end{algorithmic}
\end{algorithm}
\begin{algorithm}[tb!]
\caption{FO-US-HE}
\label{alg:FO-US-HE}
\begin{algorithmic}[1]
\REQUIRE $\bZ_P, \bZ_S, r$
\ENSURE Decision statistic of FO-US-HE
\STATE Compute $\bS_S^{-1/2} = (\bZ_S \bZ_S^\dag)^{-1/2}$
\STATE Compute $\bT_{P}=\bS_S^{-1/2}\bZ_P\bZ_P^\dag\bS_S^{-1/2}$
\STATE Compute the eigenvalues $\sigma^2_1\leq\ldots\leq\sigma^2_N$ of $\bT_{P}$
\STATE If $\min(N,K_P)>r+1$ go to step 5 else go to step 6
\STATE \textbf{Output:} $\prod\limits_{i=N-r+1}^{N} \left( 1 + \sigma^2_i \right)$
\STATE \textbf{Output:} ${\det}\left( \bI_N + \bS_S^{-1/2} \bZ_P \bZ_P^{\dag} \bS_S^{-1/2}\right)$
 \end{algorithmic}
\end{algorithm}
\begin{algorithm}[tb!]
\caption{FO-US-PHE}
\label{alg:FO-US-PHE}
\begin{algorithmic}[1]
\REQUIRE $\bZ_P, \bZ_S, r$
\ENSURE Decision statistic of FO-US-PHE
\STATE If 
$\min(K_P,N) \geq r+1$ and
$\min(K_P,N) > \frac{NK_P}{K} +r$ go to step 2 else end
\STATE Compute $\bS_S^{-1/2} = (\bZ_S \bZ_S^\dag)^{-1/2}$
\STATE Compute $\bT_{P}=\bS_S^{-1/2}\bZ_P\bZ_P^\dag\bS_S^{-1/2}$
\STATE Compute the eigenvalues $\sigma^2_1\leq\ldots\leq\sigma^2_N$ of $\bT_{P}$
\STATE Compute $\widehat{\gamma}_0$ using Corollary 2 of \cite{SubSpFrame_PART_I}
\STATE Compute $\widehat{\gamma}_1$ using Corollary 1 of \cite{SubSpFrame_PART_I}
\STATE \textbf{Output:} $\frac{\widehat{\gamma}_0^{N\left(1-\frac{K_P}{K}\right)} \prod\limits_{i=1}^{N} \left( \frac{1}{\widehat{\gamma}_0} + \sigma^2_i  \right)}{\widehat{\gamma}_1^{N\left(1-\frac{K_P}{K}\right)-r} \prod\limits_{i=1}^{N-r} \left( \frac{1}{\widehat{\gamma}_1} + \sigma^2_i  \right)}$
\end{algorithmic}
\end{algorithm}
\begin{algorithm}[tb!]
\caption{SO-KS-HE}
\label{alg:SO-KS-HE}
\begin{algorithmic}[1]
\REQUIRE $\bZ_P, \bZ_S, \bH$
\ENSURE Decision statistic of SO-KS-HE
\STATE Compute $L_0(\widehat{\bR}_0,  1; \bZ)$ as the logarithm of (5), with $\gamma=1$, in \cite{SubSpFrame_PART_I} 
\STATE Compute $\widehat{\bR}_{xx}$ and $\widehat{\bR}_1$  using Algorithm \ref{alg:alternating-SO-KS-HE}
\STATE Compute $L_1(\widehat{\bR}_1, \widehat{{\bR}}_{xx},\bH, 1; \bZ)$ 
\STATE \textbf{Output:} $L_1(\widehat{\bR}_1, \widehat{{\bR}}_{xx},\bH, 1; \bZ)-L_0(\widehat{\bR}_0,  1; \bZ)$
\end{algorithmic}
\end{algorithm}
\begin{algorithm}[tb!]
\caption{SO-KS-PHE}
\label{alg:SO-KS-PHE}
\begin{algorithmic}[1]
\REQUIRE $\bZ_P, \bZ_S, \bH$
\ENSURE Decision statistic of SO-KS-PHE
\STATE Compute $\bS_S = \bZ_S \bZ_S^\dag$
\STATE Compute $\bM_0=\bZ_P^{\dag}\bS_S^{-1} \bZ_P$
\STATE Compute $\widehat{\gamma}_0$, using Theorem 1 of \cite{SubSpFrame_PART_I}
\STATE Compute $L_0(\widehat{\bR}_0,  \widehat{\gamma}_0; \bZ)$ using the logarithm of (5) in \cite{SubSpFrame_PART_I} 
\STATE Compute $\widehat{\gamma}_1$, $\widehat{\bR}_{xx}$, and $\widehat{\bR}_1$  by means of Algorithm
\ref{alg:alternating-SO-KS-PHE} 
\STATE Compute $L_1(\widehat{\bR}_1, \widehat{{\bR}}_{xx},\bH, \widehat{\gamma}_1; \bZ)$
\STATE \textbf{Output:} $L_1(\widehat{\bR}_1, \widehat{{\bR}}_{xx},\bH, \widehat{\gamma}_1; \bZ)-L_0(\widehat{\bR}_0,  \widehat{\gamma}_0; \bZ)$
\end{algorithmic}
\end{algorithm}
\begin{algorithm}[tb!]
\caption{SO-US-HE}
\label{alg:SO-US-HE}
\begin{algorithmic}[1]
\REQUIRE $\bZ_P, \bZ_S, r$ 
\ENSURE Decision statistic of SO-US-HE
\STATE Compute $L_0(\widehat{\bR}_0,1;\bZ)$ given by the logarithm of (5) in \cite{SubSpFrame_PART_I} with $\gamma=1$
\STATE Compute $L_1(\widehat{\bR}_1, \widehat{\tilde{\bR}}_{xx}, 1; \bZ)$ exploiting Theorem 3 of \cite{SubSpFrame_PART_I} with $\gamma=1$
\STATE \textbf{Output:} $L_1(\widehat{\bR}_1, \widehat{\tilde{\bR}}_{xx}, 1; \bZ) - L_0(\widehat{\bR}_0,  1; \bZ)$
\end{algorithmic}
\end{algorithm}
\begin{algorithm}[tb!]
\caption{SO-US-PHE}
\label{alg:SO-US-PHE}
\begin{algorithmic}[1]
\REQUIRE  $\bZ_P, \bZ_S, r$
\ENSURE Decision statistic of SO-US-PHE
\STATE Compute $\widehat{\gamma}_0$, using Theorem 1 of \cite{SubSpFrame_PART_I}
\STATE Compute $L_0(\widehat{\bR}_0,  \widehat{\gamma}_0; \bZ)$ as the logarithm of (5) in \cite{SubSpFrame_PART_I} 
\STATE Compute $L_1(\widehat{\bR}_1, \widehat{\tilde{\bR}}_{xx}, \widehat{\gamma}_1; \bZ)$ by jointly exploiting Theorems 3 and 5 of \cite{SubSpFrame_PART_I}
\STATE \textbf{Output:} $L_1(\widehat{\bR}_1, \widehat{\tilde{\bR}}_{xx}, \widehat{\gamma}_1; \bZ)-L_0(\widehat{\bR}_0,  \widehat{\gamma}_0; \bZ)$
\end{algorithmic}
\end{algorithm}

\subsection{Estimate-and-Plug Approximations}
\label{Subsec:EaP}
Let us recall that the EP detectors presented in what follows are obtained
by applying the GLRT under the perfect knowledge of the disturbance covariance matrix and replacing the latter in the 
final decision statistic with the SCM of the secondary data
denoted by $\bS_{K_S}=(1/K_S)\bZ_S\bZ_S^\dag$. Moreover, without loss of generality,
we resort to a different formulation where the scaling factor $\gamma$ is present
in the second-order characterization of the primary data. Otherwise stated, the covariance
matrix of primary data is $\gamma \bR$ whereas that of secondary data is $\bR$.

\subsubsection{First-order models}
The hypothesis test to be solved in this case is given by \eqref{FO-HT}. 
Thus, exploiting the derivations in \cite{BDMGR,BBORS2007,Scharf-Friedlander1994}, 
it is possible to prove the following results.
\begin{itemize}
\item {\bf Known subspace $\langle \bH \rangle$, known $\gamma$}: assuming $\gamma=1$,
the EP approximation to the GLRT
is
\be
\tr[\bZ_P^\dag \bS_{K_S}^{-1/2} \bP_{H_S} \bS_{K_S}^{-1/2} \bZ_P]\test\eta,
\label{eq:EP-FO-KS-HE}
\ee
where $\bP_{H_S}=\bH_S (\bH_S^\dag\bH_S)^{-1} \bH_S^\dag$ with $\bH_S=\bS_{K_S}^{-1/2} \bH$.
This detector will be referred to as the EP 
approximation to the first-order detector for a signal in a known subspace in a homogeneous environment (EP-FO-KS-HE).
\item {\bf Known subspace $\langle \bH \rangle$, unknown $\gamma$}: in this case, 
the EP approximation to the GLRT is
\be
\frac{\tr[\bZ_P^\dag \bS_{K_S}^{-1} \bZ_P]}
{\tr[\bZ_P^\dag \bS_{K_S}^{-1/2} \bP^\perp_{H_S} \bS_{K_S}^{-1/2} \bZ_P]}\test \eta,
\ee
where $\bP^\perp_{H_S}=\bI_N-\bP_{H_S}$.
This detector will be referred to as the EP approximation to the first-order detector for a signal in a known subspace in a 
partially-homogeneous environment (EP-FO-KS-PHE). 
\item {\bf Unknown subspace $\langle \bH \rangle$, known $\gamma$}:
in this case, the EP approximation to the GLRT is
\be
\sum_{i=1}^{\min\{r,K_P\}}\sigma^2_i\test \eta,
\ee
where $\sigma^2_1\geq \ldots \geq \sigma^2_{N}\geq 0$ are the eigenvalues 
of $\bS_{K_S}^{-1/2}\bZ_P\bZ_P^\dag\bS_{K_S}^{-1/2}$.
This detector will be referred to as the EP approximation to the first-order detector for a signal in an unknown subspace in a 
homogeneous environment (EP-FO-US-HE). 
\item {\bf Unknown subspace $\langle \bH \rangle$, unknown $\gamma$}: in this case, the EP approximation to the GLRT is
\be
\frac{\ds\sum_{i=1}^{\min\{r,K_P\}}\sigma^2_i}
{\tr[\bZ_P^\dag\bS_{K_S}^{-1}\bZ_P]}
\test \eta.
\ee
This detector will be referred to as the EP approximation to the first-order detector for a signal in an unknown subspace in a 
partially-homogeneous environment (EP-FO-US-PHE). 
\end{itemize}

\subsubsection{Second-order models}
The hypothesis test under consideration is now problem \eqref{SO-HT}.
As in the previous subsection, we distinguish four cases.
\begin{itemize}
\item {\bf Known subspace $\langle \bH \rangle$, known $\gamma$}: without loss of generality  $\gamma=1$
and the EP approximation to the GLRT is
\cite{Bresler,ASAP04}
\be
\tr[\bB]-K_P \sum_{i=1}^{r_B} \log(1+\widehat{\lambda}_i)-\sum_{i=1}^{r_B} \frac{\gamma_i}{1+\widehat{\lambda}_i}
\test\eta,
\ee
where $\bB=\bL^{-1}\bG^\dag \bS_{K_S}^{-1/2} \bZ_P \bZ_P ^\dag \bS_{K_S}^{-1/2}  \bG\bL^{-\dag}\in\C^{r\times r}$
with rank $r_B\leq r$, $\bL\in\C^{r\times r}$ is such that $\bL\bL^\dag=\bG^\dag\bG$, and
$\widehat{\lambda}_i=\max(\gamma_i/K_P-1,0)$, $i=1,\ldots,r_B$,
with $\gamma_i$, $i=1,\ldots,r_B$, the eigenvalues of $\bB$;
$\tr[\bB]=\tr[\bZ_P^\dag \bS_{K_S}^{-1/2} \bP_G \bS_{K_S}^{-1/2} \bZ_P]$.
This detector will be referred to as the EP approximation to the second-order detector for a signal in a known subspace in a 
homogeneous environment (EP-SO-KS-HE). 
\item {\bf Known subspace $\langle \bH \rangle$, unknown $\gamma$}:
the EP approximation to the GLR is
\cite{Bresler,ASAP04}
\begin{multline}
K_PN\log\tr[\bZ_P^\dag\bS_{K_S}^{-1}\bZ_P]-K_PN\log \widehat{\gamma}
\\
- \frac{1}{\widehat{\gamma}} \tr\left( \bZ_P^\dag \bS_{K_S}^{-1/2} \bP^\perp_{G} \bS_{K_S}^{-1/2} \bZ_P \right)
\\ 
- K_P \sum_{i=1}^{r_B} \log(1+\widehat{\delta}_i)
- \sum_{i=1}^{r_B} \frac{\gamma_i/\widehat{\gamma}}{1+\widehat{\delta}_i},
\end{multline}
where $\widehat{\delta}_i=\max(\gamma_i/(K_P\widehat{\gamma})-1,0)$, $i=1,\ldots,r_B$,
and $\widehat{\gamma}$ is the solution of
\be
-\frac{K_PN}{\gamma}+\frac{\tr\left( \bZ_P^\dag \bS_{K_S}^{-1/2} \bP^\perp_{G} \bS_{K_S}^{-1/2} \bZ_P \right)}{\gamma^2}+h(\gamma)=0
\ee
with
\be
h(\gamma)=
\begin{cases}
\frac{K_P r_B}{\gamma}, & \mbox{if } \gamma < \frac{\gamma_{r_B}}{K_P},
\\
\frac{K_P(i-1)}{\gamma}+\frac{\sum_{j=i}^{r_B}\gamma_j}{\gamma2}, & \mbox{if } {\frac{\gamma_i}{K_P} \leq \gamma < 
\frac{\gamma_{i-1}}{K_P} \atop \ i=2,\ldots,r_B},
\\
\frac{\sum_{i=1}^{r_B}\gamma_i}{\gamma^2}, & \mbox{if } \frac{\gamma_1}{K_P} \leq \gamma,
\end{cases}
\ee
that maximizes the likelihood function.
This detector will be referred to as the EP approximation to the second-order detector for a signal in a known subspace in a 
partially-homogeneous environment (EP-SO-KS-PHE). 
\item {\bf Unknown subspace $\langle \bH \rangle$, known $\gamma$}:
since $\bH$ is unknown, then $\tilde{\bR}_{xx}=\bH\bR_{xx}\bH^\dag$ 
is an unknown positive semidefinite matrix with rank less than
or equal to $r$. Thus, reasoning in terms of $\tilde{\bR}_{xx}$ and following the 
lead of \cite{9321174,8781902}, the 
EP approximation to the GLRT is
\be
\tr[\bZ_P^\dag\bS_{K_S}^{-1}\bZ_P]-K_P\sum_{i=1}^r\log(1+\widehat{q}_i)-\sum_{i=1}^N\frac{\sigma^2_i}{1+\widehat{q}_i}\test\eta,
\ee
where $\widehat{q}_i=\max(\sigma^2_i/K_P-1,0)$, $i=1,\ldots,r$, $\widehat{q}_i=0$, $i=r+1,\ldots,N$, and
$\sigma^2_i$ are sorted in descending order.
This detector will be referred to as the EP approximation to the second-order detector for a signal in an unknown subspace in a 
homogeneous environment (EP-SO-US-HE). 
\item {\bf Unknown subspace $\langle \bH \rangle$, unknown $\gamma$}:
denote by $r_0$ the rank of $\bS_{K_S}^{-1/2}\bZ_P\bZ_P^\dag\bS_{K_S}^{-1/2}$; then, if $r_0\leq r$,
the EP approximation to the GLRT is
\begin{multline}
K_PN\log\tr[\bZ_P^\dag\bS_{K_S}^{-1}\bZ_P]-K_P\sum_{i=1}^{r_0} \log(\widehat{\gamma}+\widehat{q}_i)
\\
-K_P\sum_{i=r_0+1}^N\log\widehat{\gamma}
-\sum_{i=1}^{r_0}\frac{\sigma^2_i}{\widehat{\gamma}+\widehat{q}_i}
\test\eta,
\end{multline}
where $\widehat{q}_i=\max(\sigma^2_i/K_P-\widehat{\gamma},0)$, $i=1,\ldots,r_0$, 
$\widehat{q}_i=0$, $i=r_0+1,\ldots,r$, 
and $\widehat{\gamma}$ is the
solution of the equation
\be
\begin{cases}
-\frac{K_P(N-r_0)}{\gamma}=0, \mbox{ if } \frac{\sigma^2_{r_0}}{K_P}>\gamma,
\\
-\frac{K_P(N-i+1)}{\gamma} + \frac{\sum_{j=i}^{r_0}\sigma^2_j}{\gamma^2}=0, 
\mbox{ if } {\frac{\sigma^2_{i}}{K_P}\leq\gamma <\frac{\sigma^2_{i-1}}{K_P} \atop i=2,\ldots,r_0},
\\
-\frac{K_P N}{\gamma} + \frac{\sum_{i=1}^{r_0}\sigma^2_i}{\gamma^2}=0, \mbox{ if } \frac{\sigma^2_{1}}{K_P}\leq \gamma,
\end{cases}
\ee
that maximizes the likelihood function.
On the other hand, when $r_0> r$
\begin{multline}
K_PN\log\tr[\bZ_P^\dag\bS_{K_S}^{-1}\bZ_P]-K_P\sum_{i=1}^{r} \log(\widehat{\gamma}+\widehat{q}_i)
\\
-K_P\sum_{i=r+1}^N\log\widehat{\gamma}
-\sum_{i=1}^{r}\frac{\sigma^2_i}{\widehat{\gamma}+\widehat{q}_i}
-\sum_{i=r+1}^{r_0}\frac{\sigma^2_i}{\widehat{\gamma}}
\test\eta,
\end{multline}
where $\widehat{q}_i=\max(\sigma^2_i/K_P-\widehat{\gamma},0)$, $i=1,\ldots,r$, and $\widehat{\gamma}$ is the
solution of the equation
\be
\begin{cases}
-\frac{K_P(N-r)}{\gamma}+\frac{\sum_{i=r+1}^{r_0}\sigma^2_i}{\gamma^2}=0, \mbox{ if } \frac{\sigma^2_{r}}{K_P}>\gamma,
\\
-\frac{K_P(N-i+1)}{\gamma} + \frac{\sum_{j=i}^{r_0}\sigma^2_j}{\gamma^2}=0, 
\mbox{ if } {\frac{\sigma^2_{i}}{K_P}\leq\gamma <\frac{\sigma^2_{i-1}}{K_P} \atop i=2,\ldots,r},
\\
-\frac{K_P N}{\gamma} + \frac{\sum_{i=1}^{r_0}\sigma^2_i}{\gamma^2}=0, \mbox{ if } \frac{\sigma^2_{1}}{K_P}\leq \gamma,
\end{cases}
\ee
that maximizes the likelihood function.
This detector will be referred to as the EP approximation to the second-order detector for a signal in an unknown subspace in a 
partially-homogeneous environment (EP-SO-US-PHE). 
\end{itemize}

\section{Illustrative Examples and Discussion}
\label{sec:Performance}
In this section, 
Monte Carlo (MC) counting techniques are used to evaluate 
the performances of the GLR detectors derived in 
\cite{SubSpFrame_PART_I}, and these are compared to the performances of  their EP approximations. 

The probability of detection ($P_d$) and the thresholds to guarantee a given probability of false alarm ($P_{fa}$) 
are estimated over $10^3$ and $100/P_{fa}$ independent MC trials, respectively. In
all the illustrative examples we assume $N = 16$, $r=2$, and $P_{fa}=10^{-3}$ while $K_S \in \{32, 64\}$.
The covariance matrix, $\bR$, is $\bR = \bI_N + \sigma_c^2 \bM_c$, with $\sigma_c^2$ accounting 
for a clutter-to-noise ratio of 30 dB assuming unit noise power. 
The $(i, j)$th entry of the clutter component $\bM_c$ is $\rho_c^{|i-j|}$ with $\rho_c = 0.95$. 
The value of $\gamma$ for the partially-homogeneous environment is set to 2 (3 dB).

In the simulated scenario 
the signal component in the $i$th vector $\bz_i$, $i=1,\ldots,K_P$,
is given by $\alpha_i\mathbf{v}(\phi_i)$,
with
$\mathbf{v}(\phi_i) = \frac{1}{\sqrt{N}} \left[ 1 \ e^{j\phi_i} \ \cdots \ e^{j(N-1)\phi_i } \right]^T$; the electrical angles $\phi_i$ are independent random variables uniformly distributed on $\Phi=
\left[ -\pi \beta, \pi \beta \right]$, 
where $\beta=\sin \theta$ and $\theta$ equals 
$2 \pi (2/360)$ radians (corresponding to
$2^{\circ}$). 
The interval $\Phi$ is discretized using a step of 0.02 radians. Accordingly, we choose the signal subspace 
by computing the matrix $\bR_{\beta} \in \C^{N \times N}$, 
whose $(m,n)$th entry is given by \cite{4490112}
\[
\bR_{\beta}(m,n) = 2 \beta \pi \sinc((n-m)\beta).
\]
The 
signal subspace is chosen to be $<\bU_r>$ where the matrix $\bU_r \in \C^{N \times r}$ is composed of the first $r$ columns of $\bU \in \C^{N \times N}$
that in turn consists of the normalized eigenvectors of $\bR_{\beta}$ corresponding 
to its eigenvalues sorted in descending order.

\subsection{First-order models}

In this case, we define $\bV_P= \left[\mathbf{v}(\phi_1) \cdots \mathbf{v}(\phi_{K_P})\right]$ and  set 
the magnitude of $\alpha_i$, say $|{\alpha}|$, according to
the signal-to-interference-plus-noise ratio (SINR) 
defined as
\begin{equation}
\text{SINR} = |{\alpha}|^2 \tr( \bV_P^\dag \bR^{-1} \bV_P).
\label{eq_SINR_HE}
\end{equation}
The phases of the $\alpha_i$s are independent and uniformly distributed in $[0,2\pi)$.

\begin{figure}[htbp]
\begin{center}
\begin{subfigure}{.35\textwidth}
  \caption{$P_{fa}$ versus $\sigma_c^2$.}
\includegraphics[scale=0.4]{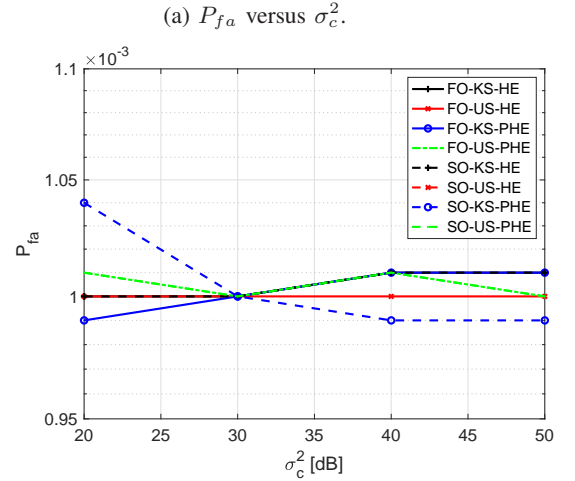}
  \label{fig:sub-a}
\end{subfigure}
\end{center}
\begin{center}
\begin{subfigure}{.35\textwidth}
  \centering
  \caption{$P_{fa}$ versus $\gamma$.}
\includegraphics[scale=0.4]{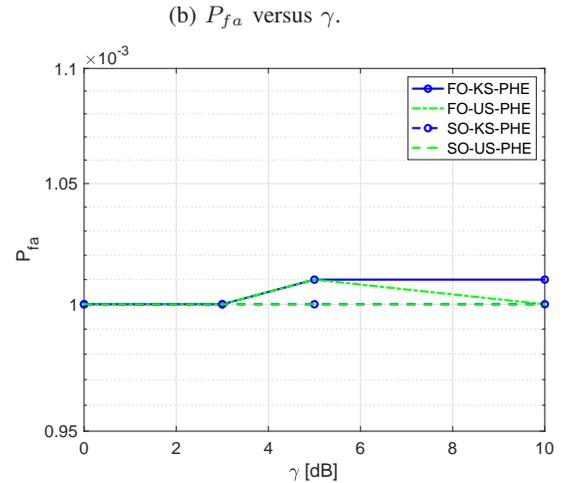}
  \label{fig:sub-b}
\end{subfigure}
\end{center}
\caption{Estimated $P_{fa}$ versus $\sigma_c^2$ (a) and $\gamma$ (b) for $N=16$, $K_P=16$, $K_S=32$, and $r=2$. 
The maximum number of iterations for the alternating procedure is 5. 
The nominal values of $\sigma_c^2$, $\gamma$, and $P_{fa}$ are $30$ dB, $3$ dB, and $10^{-3}$, respectively.}
\label{fig:CFARanalysis}
\end{figure}

The analysis starts by assessing to what extent the detection thresholds are sensitive to the variations of $\sigma_c^2$ and $\gamma$. 
The results are shown in Figure \ref{fig:CFARanalysis}, where we plot the estimated $P_{fa}$ over $100/P_{fa}$ MC trials assuming a nominal value of $10^{-3}$. These results indicate that $P_{fa}$ for all the 
derived detectors is relatively invariant to 
$\sigma_c^2$ and $\gamma$, at least for the considered parameter settings.

\begin{figure}[htbp]
\begin{center}
\includegraphics[scale=0.4]{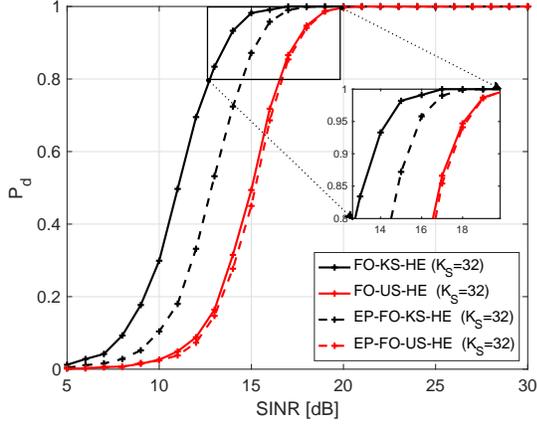}
\caption{First-order detectors for homogeneous environment: $N=16, K_P=16, r=2$, and $K_S=32$.}
\label{fig_pdet_Firstorder_HE_KS32}
\end{center}
\end{figure}
\begin{figure}[htbp]
\begin{center}
\includegraphics[scale=0.4]{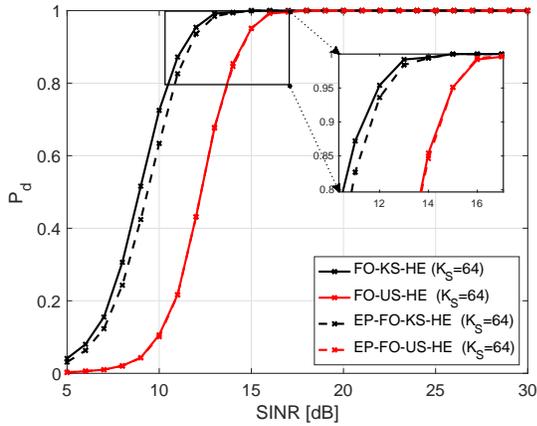}
\caption{First-order detectors for homogeneous environment: $N=16, K_P=16, r=2$, and $K_S=64$.}
\label{fig_pdet_Firstorder_HE_KS64}
\end{center}
\end{figure}

\begin{figure}[htbp]
\begin{center}
\includegraphics[scale=0.4]{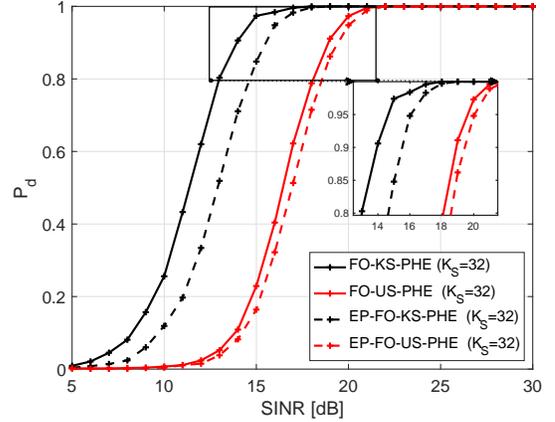}
\caption{First-order detectors for partially-homogeneous environment: $N=16, K_P=16, r=2$, and $K_S=32$.}
\label{fig_pdet_Firstorder_PHE_KS32}
\end{center}
\end{figure}

\begin{figure}[htbp]
\begin{center}
\includegraphics[scale=0.4]{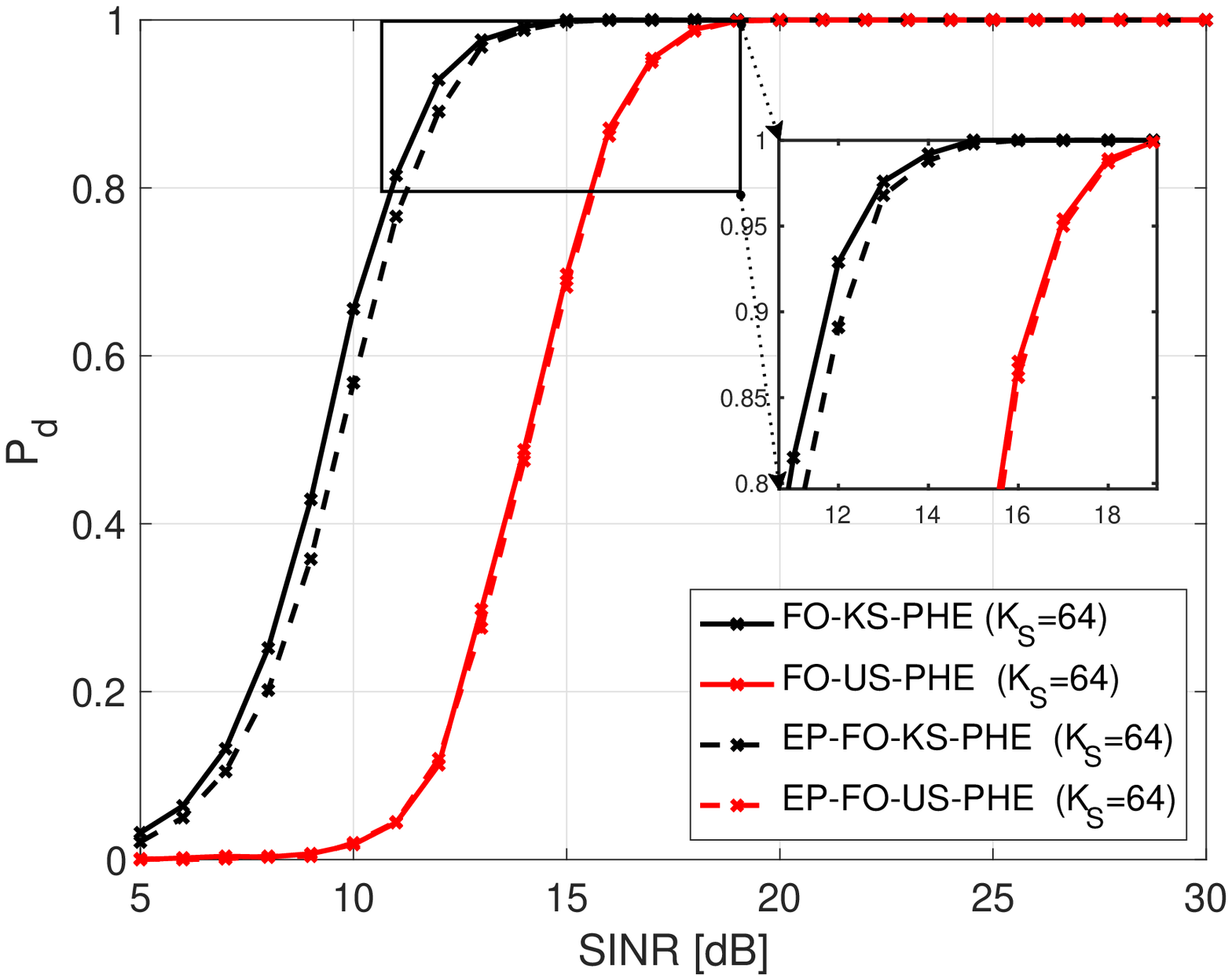}
\caption{First-order detectors for partially-homogeneous environment: $N=16, K_P=16, r=2$, and $K_S=64$.}
\label{fig_pdet_Firstorder_PHE_KS64}
\end{center}
\end{figure}

Figures \ref{fig_pdet_Firstorder_HE_KS32}-\ref{fig_pdet_Firstorder_PHE_KS64}
are plots of $P_d$ vs SINR for the first-order GLR detectors and their EP approximations. 
 Figures \ref{fig_pdet_Firstorder_HE_KS32} and
\ref{fig_pdet_Firstorder_HE_KS64}
assume a homogeneous environment
and
\ref{fig_pdet_Firstorder_PHE_KS32} 
and \ref{fig_pdet_Firstorder_PHE_KS64}
assume a partially-homogeneous environment.
The GLR detectors of \cite{SubSpFrame_PART_I} are represented by solid lines and the EP approximations 
are represented by dashed lines. Curves of detectors for a known signal subspace are black and curves of 
detectors for an unknown subspace are red.
A zoom box on high values of $P_d$ demonstrates 
the gains/losses at $P_{d}=0.9$.
Inspection of the figures shows that detectors for a known signal subspace outperform detectors for an unknown subspace, as could be expected. More importantly, GLR detectors for a known 
signal subspace outperform their EP approximations.
The GLR and EP detectors
are more or less equivalent under the assumption that the signal subspace is unknown.

To show the influence of $K_S$ on the detection performance, one can compare Figures \ref{fig_pdet_Firstorder_HE_KS32} 
and \ref{fig_pdet_Firstorder_HE_KS64} for the homogeneous environment and, similarly,
Figures \ref{fig_pdet_Firstorder_PHE_KS32} 
and \ref{fig_pdet_Firstorder_PHE_KS64} for the partially-homogeneous environment.
The comparisons highlight
the better performance obtained for the greater value of $K_S$ for all detectors, with the EP detectors filling the performance gap at $K_S=64$ due to an enhanced fidelity of the SCM estimate.
Additional numerical examples not reported here for brevity confirm the observed behavior for $r=4$.

\begin{figure}[htbp]
\label{fig_convergence}
\begin{center}
\begin{subfigure}{.35\textwidth}
  \centering
  \caption{Homogeneous Environment ($H_0$).}
\includegraphics[scale=0.345]{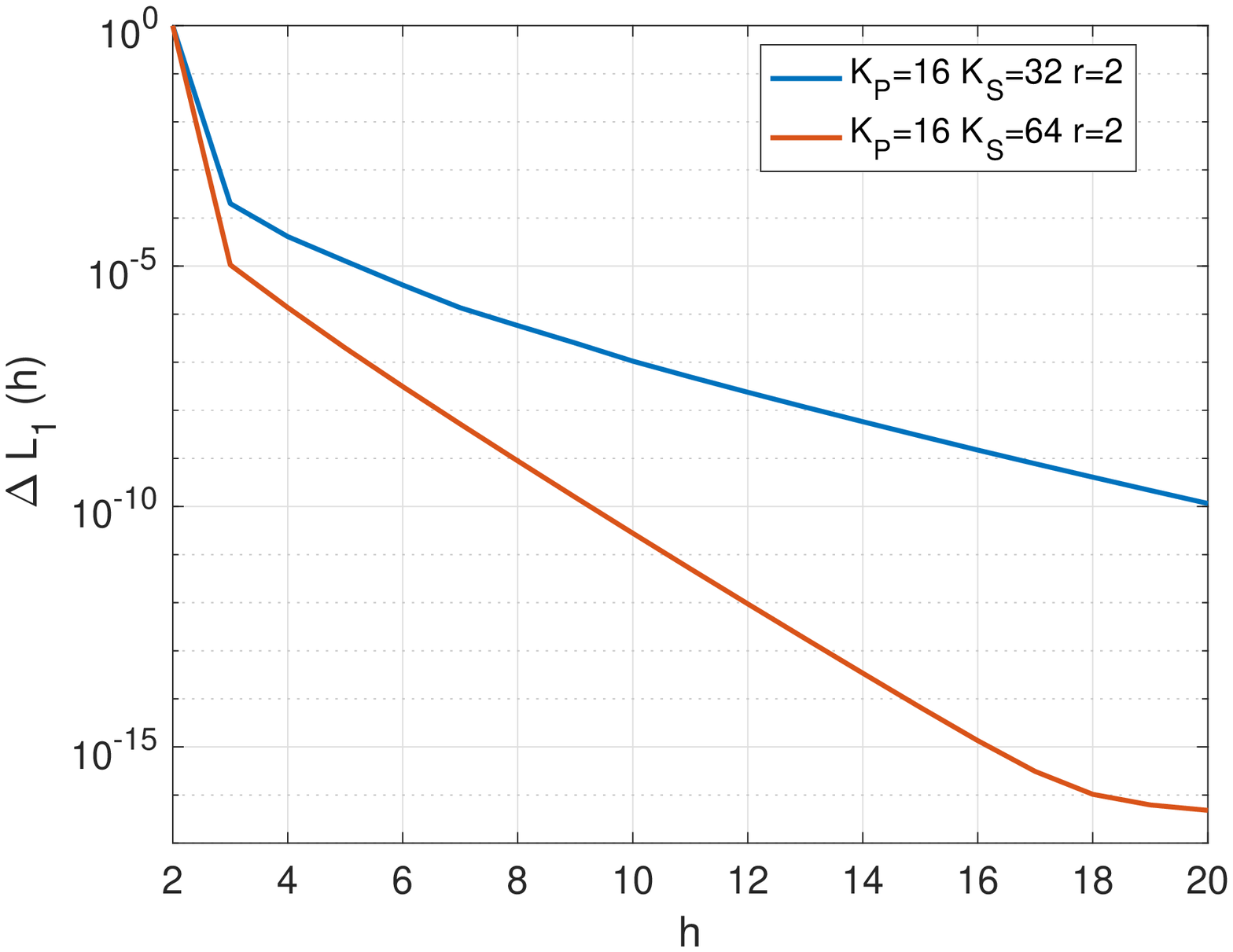}
  \label{fig:sub-a}
\end{subfigure}
\end{center}
\begin{center}
\begin{subfigure}{.35\textwidth}
  \centering
  \caption{Homogeneous Environment ($H_1$).}
\includegraphics[scale=0.345]{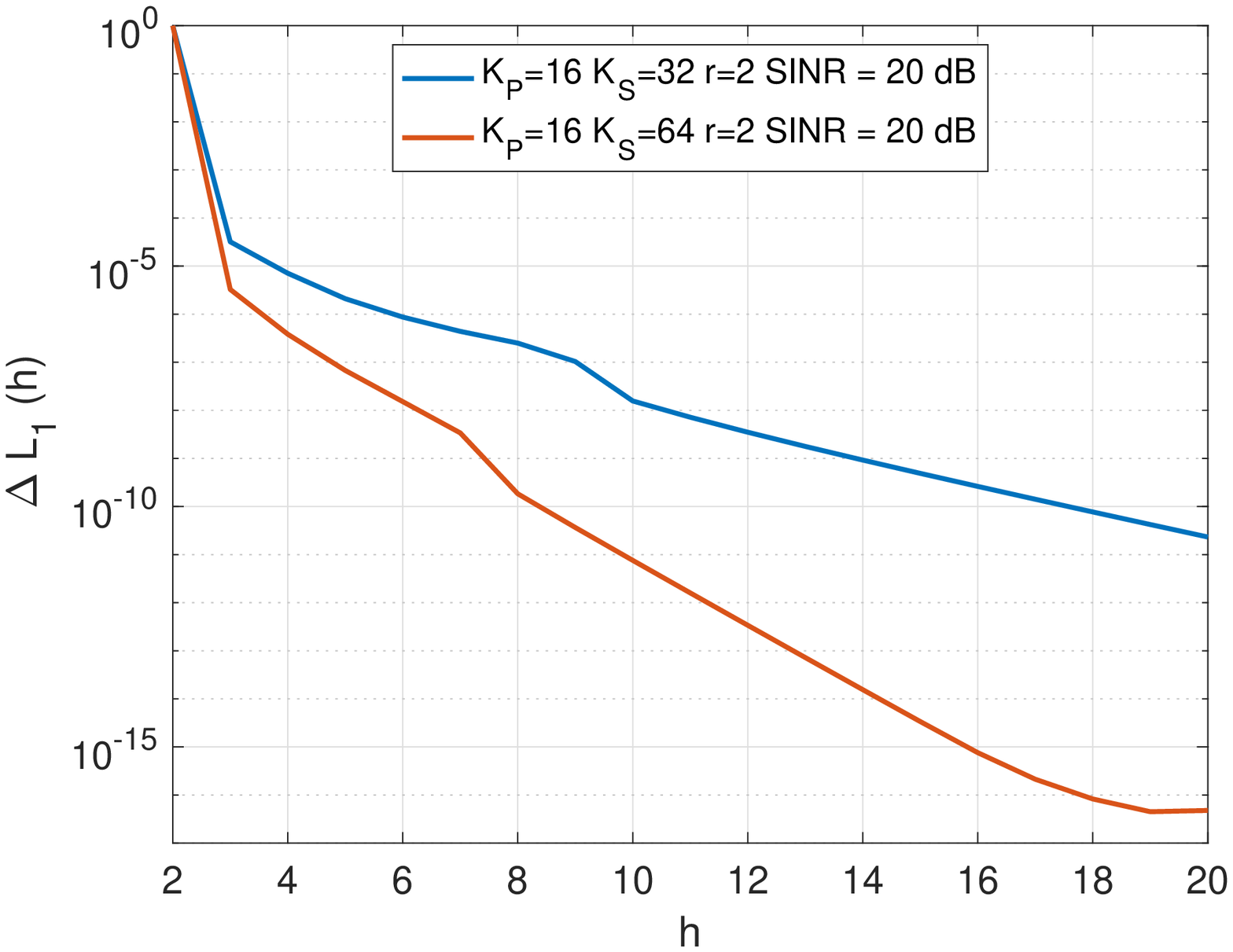}
  \label{fig:sub-b}
\end{subfigure}
\end{center}
\begin{center}
\begin{subfigure}{.35\textwidth}
  \centering
  \caption{Partially-Homogeneous Environment ($H_0$).}
\includegraphics[scale=0.345]{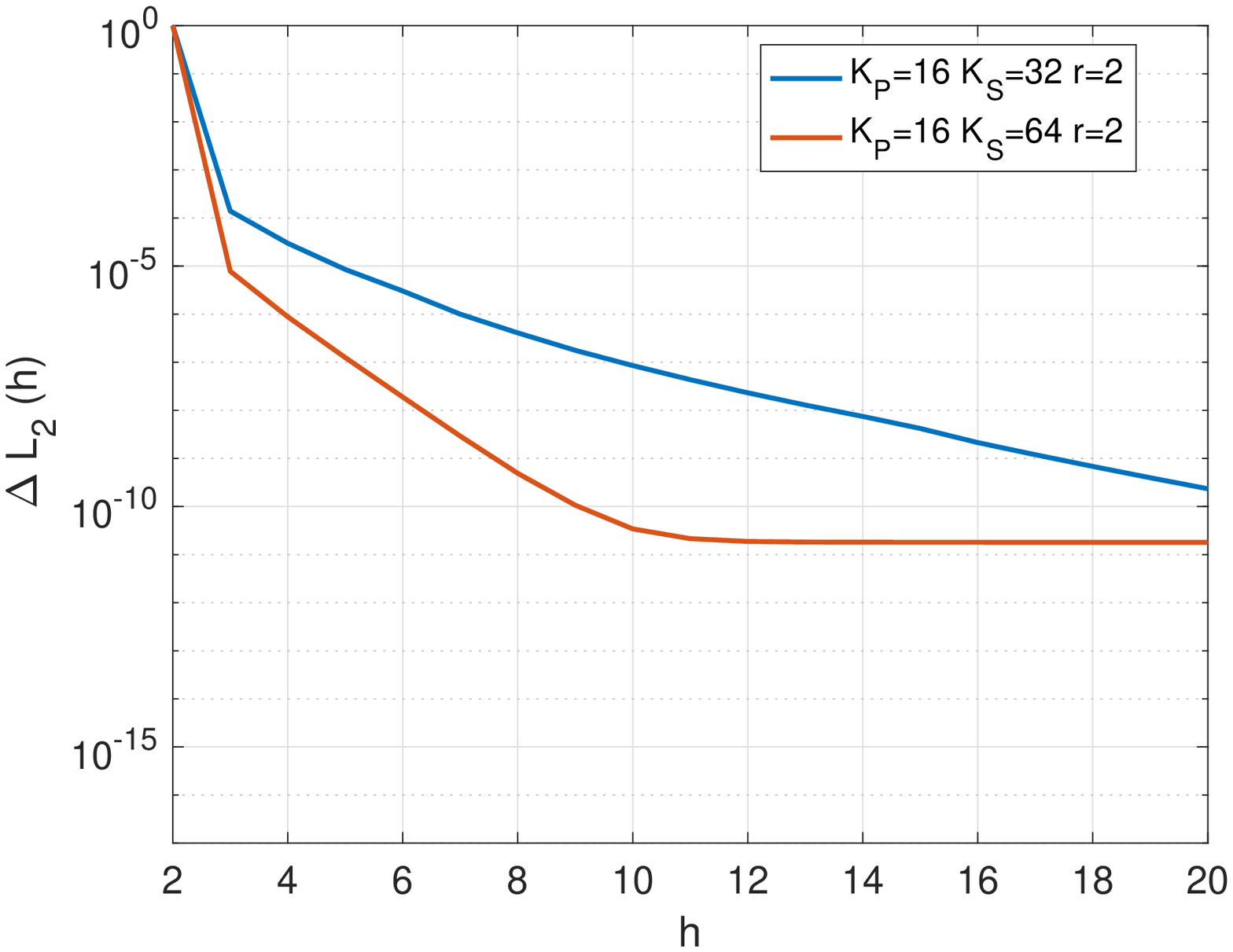}
  \label{fig:sub-c}
\end{subfigure}
\end{center}
\begin{center}
\begin{subfigure}{.35\textwidth}
  \centering
  \caption{Partially-Homogeneous Environment ($H_1$).}
\includegraphics[scale=0.345]{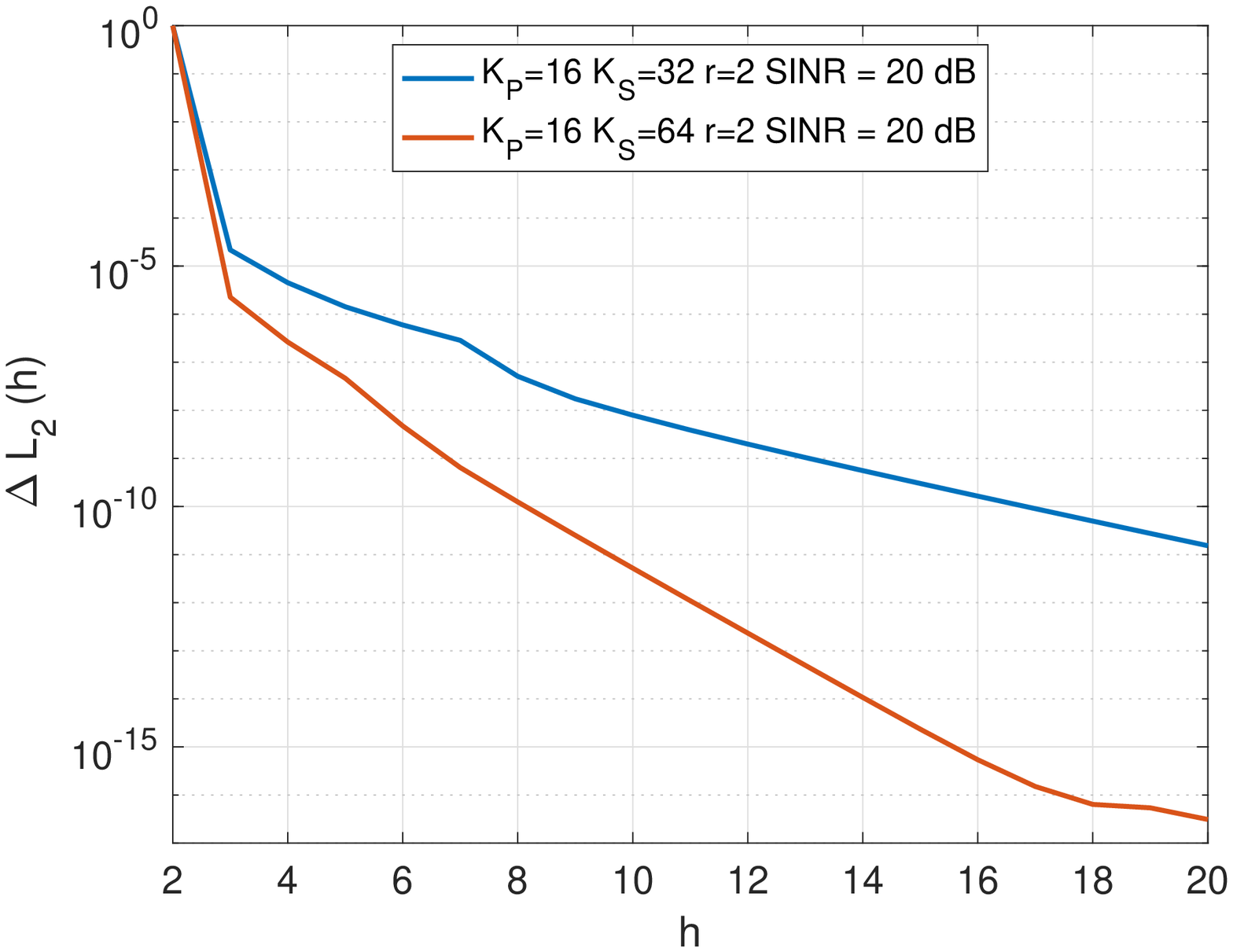}
  \label{fig:sub-d}
\end{subfigure}
\end{center}
\caption{Log-likelihood variation versus the iteration number of the alternating procedures.}
\end{figure}

\subsection{Second order models}
Under the second-order model, 
$\boldsymbol{\alpha}=[\alpha_1,\dots,\alpha_{K_P}]^T$ 
is a complex Gaussian vector with covariance matrix $\sigma_{\alpha}^2 \bI_{K_P}$, 
with $\sigma^2_{\alpha}>0$ varying according to the SINR defined in \eqref{eq_SINR_HE} 
with $\sigma_{\alpha}^2$ replacing $|\alpha|^2$.

As a preliminary step, we analyze the proposed alternating procedures for iterations $h$, ranging from 2 to 20. To this end, we plot the average values of $\Delta L_i$, $i=1,2$,
over 100 MC trials versus $h$, in Figures \ref{fig:sub-a}-\ref{fig:sub-d}, for both the
homogeneous and the partially-homogeneous environments and simulating the null and the alternative hypotheses. 
All the parameter values used for this analysis are shown in the figures; under $H_1$ the SINR value is set to $20$ dB. 
It turns out that, for the considered parameters, 5 iterations are sufficient 
to achieve a relative variation approximately lower than $10^{-5}$ and this value is also used in what follows.

\begin{figure}[h!]
\begin{center}
\includegraphics[scale=0.4]{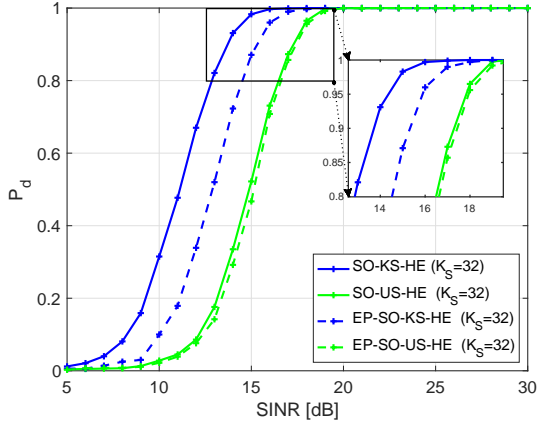}
\caption{Second-order detectors for homogeneous environment: $N=16, K_P=16, r=2$, and $K_S=32$.}
\label{fig_pdet_Secondorder_HE_KS32}
\end{center}
\end{figure}

\begin{figure}[htbp]
\begin{center}
\includegraphics[scale=0.4]{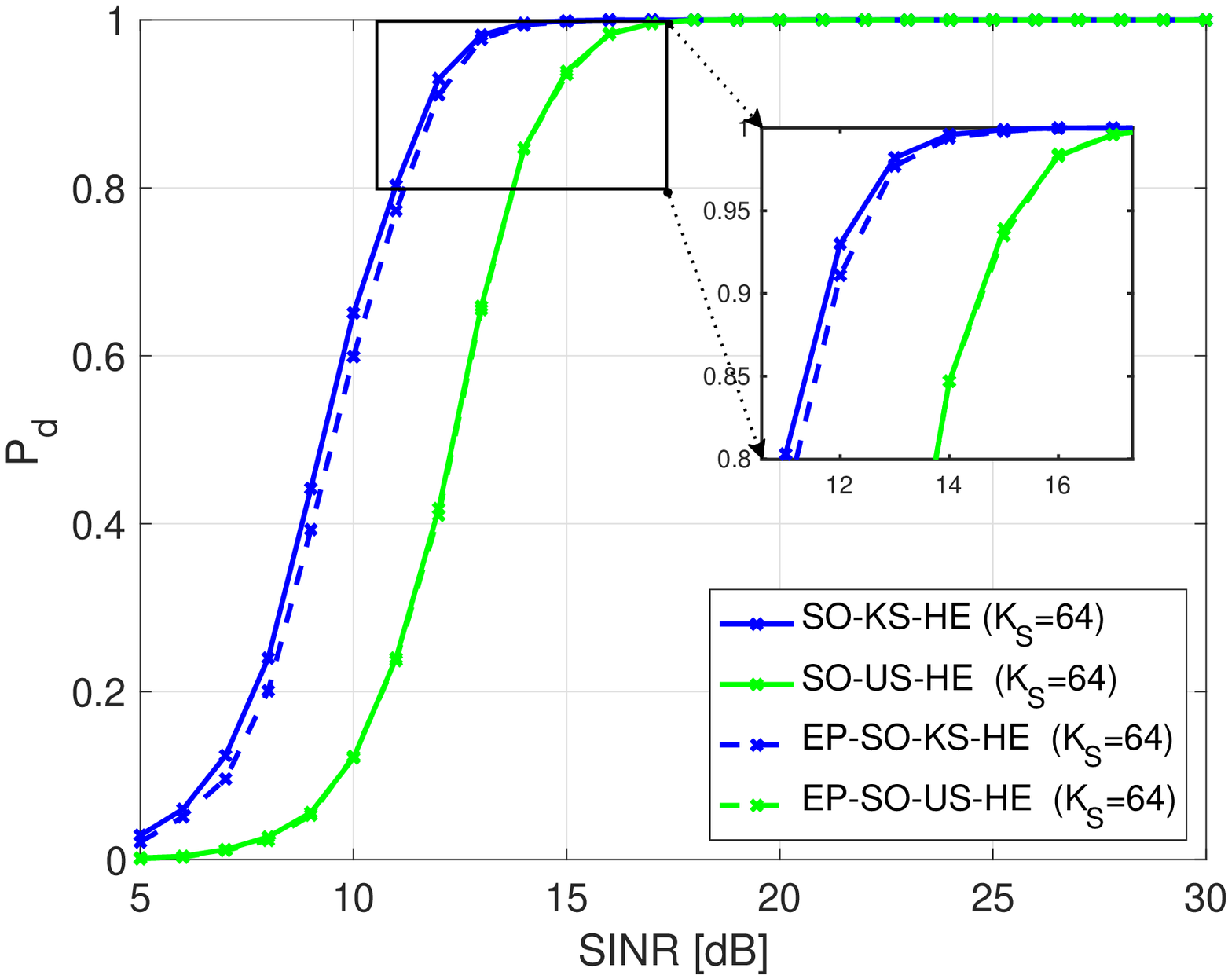}
\caption{Second-order detectors for homogeneous environment: $N=16, K_P=16, r=2$, and $K_S=64$.}
\label{fig_pdet_Secondorder_HE_KS64}
\end{center}
\end{figure}

\begin{figure}[htbp]
\begin{center}
\includegraphics[scale=0.4]{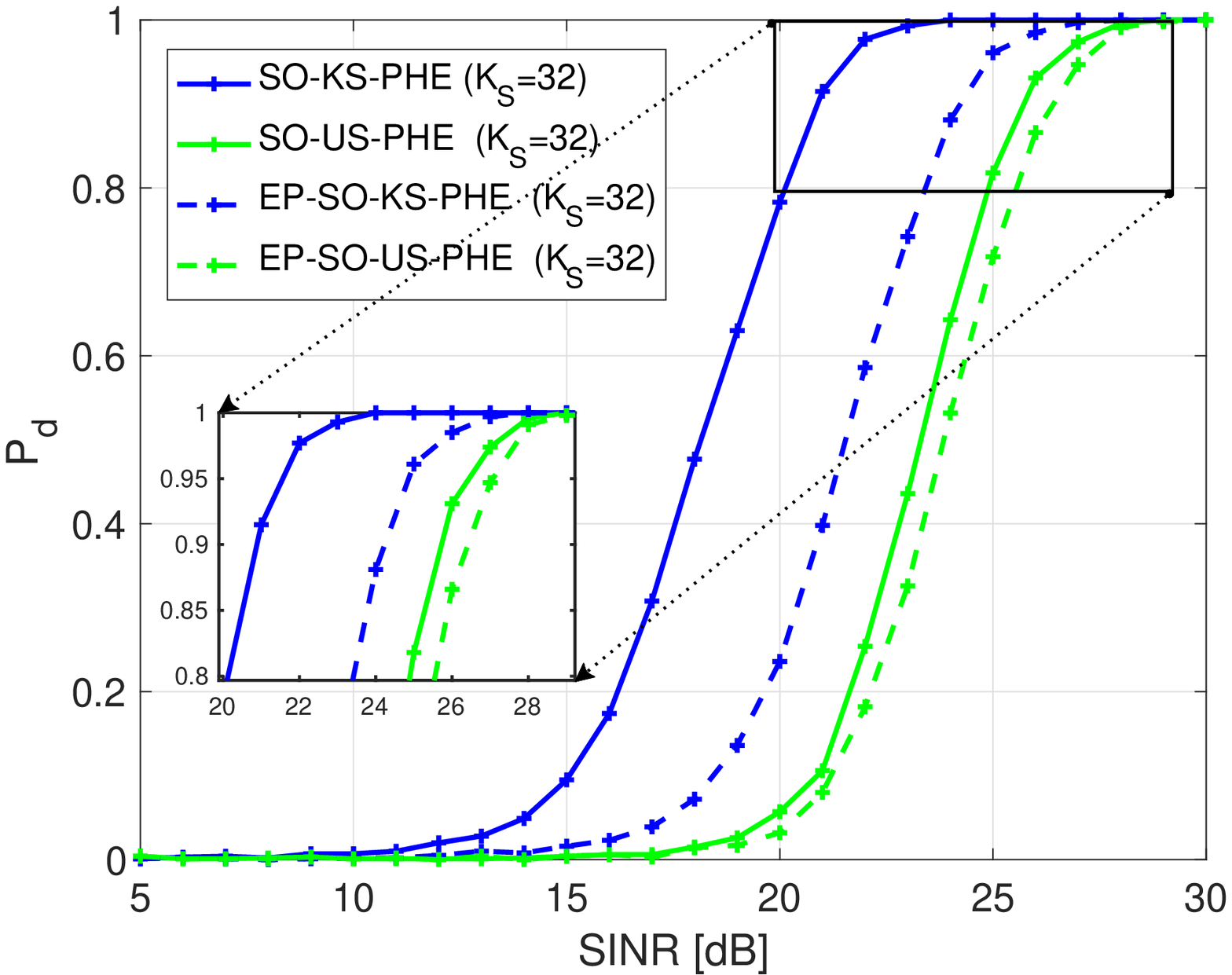}
\caption{Second-order detectors for partially-homogeneous environment: $N=16, K_P=16, r=2$, and $K_S=32$.}
\label{fig_pdet_Secondorder_PHE_KS32}
\end{center}
\end{figure}

\begin{figure}[htbp]
\begin{center}
\includegraphics[scale=0.4]{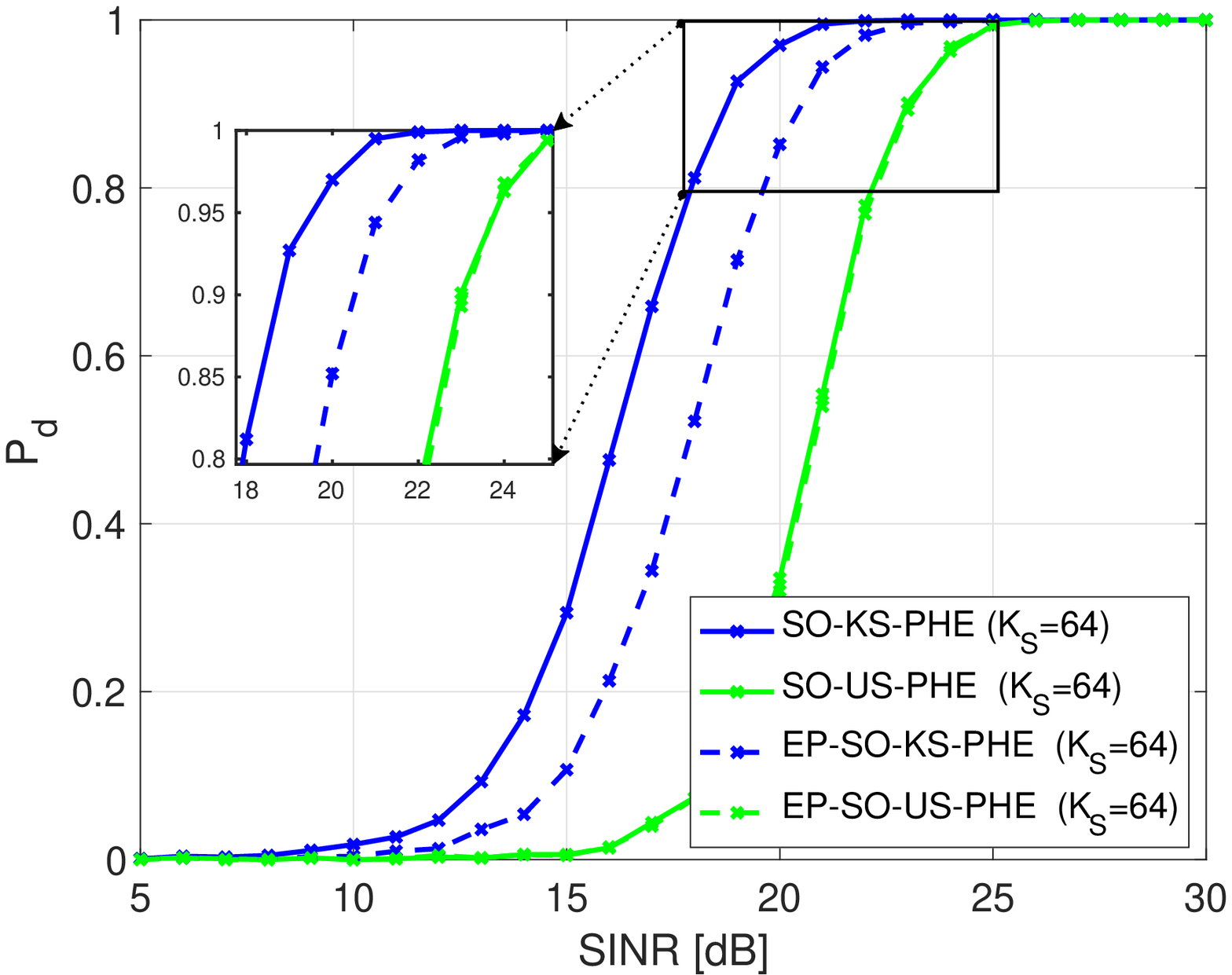}
\caption{Second-order detectors for partially-homogeneous environment: $N=16, K_P=16, r=2$, and $K_S=64$.}
\label{fig_pdet_Secondorder_PHE_KS64}
\end{center}
\end{figure}

Figures \ref{fig_pdet_Secondorder_HE_KS32}-\ref{fig_pdet_Secondorder_PHE_KS64}
are plots of  $P_d$ vs SINR for the second-order GLR detectors and their EP approximations. 
 Figures \ref{fig_pdet_Secondorder_HE_KS32} and
\ref{fig_pdet_Secondorder_HE_KS64}
assume a homogeneous environment and
\ref{fig_pdet_Secondorder_PHE_KS32} 
and \ref{fig_pdet_Secondorder_PHE_KS64}
assume a partially-homogeneous environment.
The GLR detectors proposed in \cite{SubSpFrame_PART_I} are represented by solid lines and the EP 
approximations are represented by dashed lines. Curves of detectors for a known signal subspace are blue and curves of 
detectors for an unknown subspace are green. Again
a zoom box on high values of $P_d$ is reported. 
The second-order detectors 
for a known signal subspace outperform detectors for an unknown signal subspace 
and GLR detectors for a known signal subspace are better than the corresponding EP 
detectors for $K_S=32$.  However, this time the gain of the GLR detector over 
the corresponding EP detector is much more pronounced in a partially-homogeneous environment 
and, in the case of detectors for a known subspace, is still remarkable for $K_S=64$.

\section{Conclusions}
\label{sec:Conclusions}

In this paper, we have assessed the performance of the GLR detectors derived in the companion paper
\cite{SubSpFrame_PART_I} and compared the performance of these detectors to the performance of EP approximations.
It is worth noticing that most of the EP approximations have been derived here for the first time (at least to the best of authors' knowledge).
As in \cite{SubSpFrame_PART_I}, we have considered two operating situations: a homogeneous environment 
where training samples and testing samples  share the same statistical characterization of the interference,
and a partially-homogeneous environment where training and testing samples differ in scale.
The analysis starts by investigating to what extent the $P_{fa}$ is sensitive to variations
of the clutter parameters showing that all the GLR  detectors maintain a rather constant
false alarm rate over the considered parameter ranges. 
When the signal subspace is known, performance is better than
when it is known only by its dimension.
The GLR detectors outperform their EP approximants
in case the signal subspace is known
and the number of secondary data is not too large. 
Finally, the performance of the detectors for an unknown signal subspace are close to each other. Summarizing, the analysis
has shown that the design framework proposed in \cite{SubSpFrame_PART_I} leads to
effective solutions for signals with inherent uncertainty that, for a specific
radar application, can be related to the angles of arrival, Doppler frequency, and/or 
phase/amplitude calibration errors.
Future research might analyze these detectors on real data and under a mismatch
between the actual and the nominal signal subspace. 

\vspace{0.5 cm}


\bibliographystyle{IEEEtran}
\bibliography{group_bib}

\begin{thebibliography}{10}
\providecommand{\url}[1]{#1}
\csname url@samestyle\endcsname
\providecommand{\newblock}{\relax}
\providecommand{\bibinfo}[2]{#2}
\providecommand{\BIBentrySTDinterwordspacing}{\spaceskip=0pt\relax}
\providecommand{\BIBentryALTinterwordstretchfactor}{4}
\providecommand{\BIBentryALTinterwordspacing}{\spaceskip=\fontdimen2\font plus
\BIBentryALTinterwordstretchfactor\fontdimen3\font minus
  \fontdimen4\font\relax}
\providecommand{\BIBforeignlanguage}[2]{{%
\expandafter\ifx\csname l@#1\endcsname\relax
\typeout{** WARNING: IEEEtran.bst: No hyphenation pattern has been}%
\typeout{** loaded for the language `#1'. Using the pattern for}%
\typeout{** the default language instead.}%
\else
\language=\csname l@#1\endcsname
\fi
#2}}
\providecommand{\BIBdecl}{\relax}
\BIBdecl

\bibitem{SubSpFrame_PART_I}
D.~{Orlando}, G.~{Ricci}, and L.~L. {Scharf}, ``{A Unified Theory of Adaptive
  Subspace Detection - Part I: Detector Designs},'' \emph{IEEE Transactions on
  Signal Processing}, 2021, under review.

\bibitem{BDMGR}
F.~Bandiera, A.~De~Maio, A.~S. Greco, and G.~Ricci, ``{Adaptive Radar Detection
  of Distributed Targets in Homogeneous and Partially Homogeneous Noise Plus
  Subspace Interference},'' \emph{IEEE Transactions on Signal Processing},
  vol.~55, no.~4, pp. 1223--1237, 2007.

\bibitem{BBORS2007}
F.~{Bandiera}, O.~{Besson}, D.~{Orlando}, G.~{Ricci}, and L.~L. {Scharf},
  ``{GLRT}-based direction detectors in homogeneous noise and subspace
  interference,'' \emph{IEEE Transactions on Signal Processing}, vol.~55,
  no.~6, pp. 2386--2394, 2007.

\bibitem{Scharf-Friedlander1994}
L.~L. {Scharf} and B.~{Friedlander}, ``{Matched subspace detectors},''
  \emph{IEEE Transactions on Signal Processing}, vol.~42, no.~8, pp.
  2146--2157, 1994.

\bibitem{scharfRevewSubspace}
L.~Scharf, S.~Kraut, and M.~McCloud, ``A review of matched and adaptive
  subspace detectors,'' in \emph{Proceedings of the IEEE 2000 Adaptive Systems
  for Signal Processing, Communications, and Control Symposium (Cat.
  No.00EX373)}, 2000, pp. 82--86.

\bibitem{hyperSubspace}
Y.~Hou, W.~Zhu, E.~Wang, and Y.~Zhang, ``{A Hyperspectral Subspace Target
  Detection Method Based on AMUSE},'' \emph{International Journal of Pattern
  Recognition and Artificial Intelligence}, vol.~33, no.~12, 2019.

\bibitem{8850120}
N.~Acito, M.~Moscadelli, M.~Diani, and G.~Corsini, ``Subspace-based target
  detection in {LWIR} hyperspectral imaging,'' \emph{IEEE Geoscience and Remote
  Sensing Letters}, vol.~17, no.~6, pp. 1047--1051, 2020.

\bibitem{9387095}
C.-I. Chang, H.~Cao, and M.~Song, ``Orthogonal subspace projection target
  detector for hyperspectral anomaly detection,'' \emph{IEEE Journal of
  Selected Topics in Applied Earth Observations and Remote Sensing}, vol.~14,
  pp. 4915--4932, 2021.

\bibitem{BCCRV}
O.~Besson, A.~Coluccia, E.~Chaumette, G.~Ricci, and F.~Vincent, ``Generalized
  likelihood ratio test for detection of gaussian rank-one signals in gaussian
  noise with unknown statistics,'' \emph{IEEE Transactions on Signal
  Processing}, vol.~65, no.~4, pp. 1082--1092, February 15 2017.

\bibitem{Kelly-Forsythe}
E.~J. Kelly and K.~Forsythe, ``{Adaptive Detection and Parameter Estimation for
  Multidimensional Signal Models},'' Lincoln Lab, MIT, Lexington, US, Technical
  Report 848, 1989.

\bibitem{Kraut-Scharf1999}
S.~{Kraut} and L.~L. {Scharf}, ``The {CFAR} adaptive subspace detector is a
  scale-invariant {GLRT},'' \emph{IEEE Transactions on Signal Processing},
  vol.~47, no.~9, pp. 2538--2541, 1999.

\bibitem{subspaceNonGaussian}
M.~Desai and R.~Mangoubi, ``{Robust Gaussian and non-Gaussian matched subspace
  detection},'' \emph{IEEE Transactions on Signal Processing}, vol.~51, no.~12,
  pp. 3115--3127, 2003.

\bibitem{Gini-Farina2002}
F.~{Gini} and A.~{Farina}, ``Vector subspace detection in compound-gaussian
  clutter. {Part I}: survey and new results,'' \emph{IEEE Transactions on
  Aerospace and Electronic Systems}, vol.~38, no.~4, pp. 1295--1311, 2002.

\bibitem{ASAP04}
G.~Ricci and L.~Scharf, ``Adaptive radar detection of extended {Gaussian}
  targets,'' in \emph{The Twelfth Annual Workshop on Adaptive Sensor Array
  Processing ASAP 2004}.\hskip 1em plus 0.5em minus 0.4em\relax Lincoln
  Laboratory, Massachusetts Institute of Technology, Lexington, Massachusetts
  (USA), 16-18 March 2004.

\bibitem{Robey}
F.~Robey, D.~Fuhrmann, E.~Kelly, and R.~Nitzberg, ``{A CFAR adaptive matched
  filter detector},'' \emph{IEEE Transactions on Aerospace and Electronic
  Systems}, vol.~28, no.~1, pp. 208--216, 1992.

\bibitem{Stoica_alternating}
P.~Stoica and Y.~Selen, ``{Cyclic minimizers, majorization techniques, and the
  expectation-maximization algorithm: a refresher},'' \emph{IEEE Signal
  Processing Magazine}, vol.~21, no.~1, pp. 112--114, 2004.

\bibitem{9309189}
S.~Han, L.~Yan, Y.~Zhang, P.~Addabbo, C.~Hao, and D.~Orlando, ``{Adaptive Radar
  Detection and Classification Algorithms for Multiple Coherent Signals},''
  \emph{IEEE Transactions on Signal Processing}, vol.~69, pp. 560--572, 2021.

\bibitem{Ward1994}
J.~Ward, ``Space-time adaptive processing for airborne radar,'' MIT Lincoln
  Laboratory, Tech. Rep. 1015, 1994.

\bibitem{Bresler}
Y.~{Bresler}, ``{Maximum likelihood estimation of a linearly structured
  covariance with application to antenna array processing},'' in \emph{Fourth
  Annual ASSP Workshop on Spectrum Estimation and Modeling}, Minneapolis, MN
  (USA), August 3-5 1988, pp. 172--175.

\bibitem{BOR_MC2009}
F.~{Bandiera}, D.~{Orlando}, and G.~{Ricci}, \emph{{Advanced Radar Detection
  Schemes Under Mismatched Signal Models}}.\hskip 1em plus 0.5em minus
  0.4em\relax Synthesis Lectures on Signal Processing, Morgan \& Claypool
  Publishers, 2009.

\bibitem{9321174}
P.~Addabbo, S.~Han, D.~Orlando, and G.~Ricci, ``{Learning Strategies for Radar
  Clutter Classification},'' \emph{IEEE Transactions on Signal Processing},
  vol.~69, pp. 1070--1082, 2021.

\bibitem{8781902}
L.~Yan, P.~Addabbo, C.~Hao, D.~Orlando, and A.~Farina, ``{New ECCM Techniques
  Against Noiselike and/or Coherent Interferers},'' \emph{IEEE Transactions on
  Aerospace and Electronic Systems}, vol.~56, no.~2, pp. 1172--1188, 2020.

\bibitem{4490112}
A.~Pezeshki, B.~D. Van~Veen, L.~L. Scharf, H.~Cox, and M.~Lundberg~Nordenvaad,
  ``Eigenvalue beamforming using a multirank mvdr beamformer and subspace
  selection,'' \emph{IEEE Transactions on Signal Processing}, vol.~56, no.~5,
  pp. 1954--1967, 2008.

\end{thebibliography}

\end{document}